\documentclass[namedreferences]{solarphysics}

\usepackage[optionalrh,showbiblabels]{spr-sola-addons} 
\usepackage{graphicx}        
\usepackage{amssymb}        
\usepackage{color}           
\usepackage{breakurl}        
\usepackage{soul}

\newcommand{\curl}{ {\bf \nabla} \times}

\chardef\us=`\_

\usepackage{xcolor} 

\begin{document}

\begin{article}
\begin{opening}

\title{An MHD Study of Large-Amplitude Oscillations in Solar Filaments \\}

\author[addressref={aff1},corref,email={ernesto.zurbriggen@craam.mackenzie.br}]{\inits{E.}\fnm{Ernesto}~\lnm{Zurbriggen}}
\author[addressref={aff2,aff3},email={}]{\inits{M.}\fnm{Mariana}~\lnm{C\'ecere}}
\author[addressref={aff6},corref,email={}]{\inits{M.V.}\fnm{Mar\'ia Valeria}~\lnm{Sieyra}}
\author[addressref={aff4,aff5},corref]{\inits{G.}\fnm{Gustavo}~\lnm{Krause}}
\author[addressref={aff2},corref]{\inits{A.}\fnm{Andrea}~\lnm{Costa}}
\author[addressref={aff1},corref]{\inits{C.}\fnm{C. Guillermo}~\lnm{Gim\'enez de Castro}}

\address[id=aff1]{Universidade Presbiteriana Mackenzie, Centro de R\'adio Astronomia e Astrof\'isica Mackenzie (CRAAM), SP, S\~ao Paulo, Brasil.}
\address[id=aff2]{Consejo Nacional de Investigaciones Cient\'ificas y T\'ecnicas (CONICET), Instituto de Astronom\'ia Te\'orica y Experimental (IATE), C\'ordoba, Argentina.}
\address[id=aff3]{Universidad Nacional de C\'ordoba (UNC), Observatorio Astron\'omico de C\'ordoba (OAC), C\'ordoba, Argentina.}
\address[id=aff6]{Centre for mathematical Plasma Astrophysics, Department of Mathematics, KU Leuven, Celestijnenlaan 200B, B-3001 Leuven, Belgium}
\address[id=aff4]{Consejo Nacional de Investigaciones Cient\'ificas y T\'ecnicas, Instituto de Estudios Avanzados en Ingenier\'ia y Tecnolog\'ia (IDIT), C\'ordoba, Argentina.}
\address[id=aff5]{Universidad Nacional de C\'ordoba , Facultad de Ciencias Exactas, F\'isicas y Naturales, C\'ordoba, Argentina.}

\runningauthor{Zurbriggen et al.}
\runningtitle{Large-Amplitude Oscillations in Solar Filaments}

\begin{abstract}   
Quiescent filaments are usually affected by internal and/or external perturbations triggering oscillations of different kinds. In particular, external large-scale coronal waves can perturb remote quiescent filaments leading to large-amplitude oscillations. Observational reports have indicated that the activation time of oscillations coincides with the passage of a large-scale coronal wavefront through the filament, although the disturbing wave is not always easily detected. Aiming to contribute to understand how -and to what extent- coronal waves are able to excite filament oscillations, here we modelled with 2.5 MHD simulations a filament floating in a gravitationally stratified corona disturbed by a  coronal shock wave. This simplified scenario results in a two-coupled oscillation pattern of the filament which is damped in a few cycles, enabling a detailed analysis. A parametric study was accomplished varying parameters of the scenario such as height, size and mass of the filament. An oscillatory analysis reveals a general tendency where  periods of oscillations, amplitudes and damping times increase with height, whereas  filaments of larger radius exhibit shorter periods and smaller amplitudes. The calculation of forces exerted on the filament shows that the main restoring force is the magnetic tension. 
\end{abstract}

\end{opening}

\section{Introduction} 
\label{S-Introduction} 

Prominence seismology is a powerful diagnostic tool that combines the observation of prominence oscillations with theoretical models to infer coronal plasma parameters and to analyse the dynamical response associated with restoring forces involved in the motion \citep[see e.g.,][]{2009tripathiSSRv149,2018arregieLRSP15}. In particular, large-amplitude oscillations of prominences can be classified into longitudinal oscillations, with the motion parallel to the prominence's axis, and transverse oscillations, moving perpendicular to the axis direction. Transverse oscillations can also be divided into horizontal and vertical oscillations, and have been related to the winking filament phenomenon observed at H$\alpha$ \citep{1966AJ.....71..197R}. Ranges of periods, velocity amplitudes and damping times of 11--29~min, 6--41~km s$^{-1}$ and 25--180~min \citep{2014shenApJ795}, respectively, have been reported for transverse oscillations. Whereas longitudinal oscillations are associated with larger periods, velocity amplitudes and damping times in the ranges of 44--160~min, 30--100~km s$^{-1}$ and 115--600~min, respectively. In this work we will use equivalently the words prominence and filament.

The determination of driving mechanisms for large-amplitude oscillations are still subject to research. Nevertheless, several studies have suggested that the triggering disturbances are fast MHD shocks, such as chromospheric Moreton waves \citep{2002PASJ...54..481E,2013franchileAA552} and coronal EUV waves produced by distant flares and/or coronal mass ejections \citep{2004ApJ...608.1124O,2006A&A...449L..17I,2012asaiApJ745,2012ApJ...754....7S,2014shenApJ786,2014shenApJ795}. Subflares or jets are also able to produce large-amplitude oscillations \citep{2003ApJ...584L.103J,2021lunaApJ912}. \citet{2013franchileAA552} pointed out that the activation time of two faraway winking filaments coincided with the passage through them of a large-scale wavefront, not detected at coronal heights. Also, \cite{2014shenApJ786} analysing four winking filaments, excited by a weak coronal EUV wave coming from an X-flare region, suggested that the oscillation parameters resulted from the normal mode excitation of each filament triggered by a single disturber. Previously, \cite{1966AJ.....71..197R} proposed that the oscillation parameters depend on the intrinsic properties of the prominence rather than on the external disturbance features. In fact, if the external driver is a pulse, i.e. not a forced oscillation mechanism, the system should respond to the perturbation with its natural frequency. Moreover, considering the fluctuation-dissipation theorem \citep{1966RPPh...29..255K}, small amplitude oscillations will be spontaneously excited at the natural frequencies of the system. 
 
Several observational studies of the properties of filament oscillations contributed to understand the physical mechanisms involved in the motion. For instance, \cite{2014shenApJ795}  suggested that a prominence oscillates like a linear vertical solid body with one end tied on the solar surface. The analysis performed by \cite{2013ApJ...773..166L} and \cite{2015RAA....15.1713P} on the measurements of transverse oscillations also supported the idea that filaments oscillate as a whole. However, \cite{2011A&A...531A..53H}, by analysing this type of arched prominence oscillations, suggested that the prominence presented a global kink mode, although there were some discrepancies indicating the prominence oscillates as a collection of separated but interacting threads rather than like a rigid body. Concerning the forces, \cite{2014shenApJ795} proposed that the restoring forces of the transverse oscillations are most likely due to the coupling of gravity and magnetic tension of the supporting magnetic field. Furthermore, by analysing transverse oscillations, \cite{2008gilbertApJ685} sustained that the main restoring force is the magnetic tension. 

From another perspective, several numerical studies have been performed to contribute to the understanding of prominence seismology. Among several models available to emulate prominence dynamics, the main ones are two simplified cases associated with the magnetic structures of \citet{1957ZA.....43...36K} and \citet{1974A&A....31..189K}. In the first model the prominence is sustained by the magnetic field lines (magnetic dip), whereas in the second one it is totally contained in the interior of a closed helical magnetic structure that isolates the prominence mass. For example, performing 3D MHD simulations, \citet{2020adroverA&A633} analysed prominence oscillations generated by instantaneous velocity perturbations on a relaxed system based on the Kippenhahn-Schl\"uter model, identifying the magnetic force as the restoring process for transverse oscillations. Also, they showed that periods increase with the prominence density and width, but decreases with the magnetic field strength. On the other hand, in line with Kuperus and Raadu, \citet{2018zhouApJ856} carried out 3D ideal MHD simulations of a prominence in an initially relaxed state where the oscillations were triggered by perturbing the velocity field independently in each direction. These authors found that periods of horizontal and vertical transverse oscillations are different, and attributed this result to the shape of the prominence, which is wider in the vertical direction. In addition, they determined for transverse oscillations that the main restoring force is the magnetic tension.  \citet{2020liakhA&A637} used 2.5D simulations to model oscillations excited independently by horizontal and vertical internal perturbations, and also by an external perturbation. In the former case, considering a single prominence the authors found a weak dependence of transverse oscillation periods on the density contrast and on the shear angle, suggesting that periods are almost constant with height. For the latter case, they found that the displacement and deformation of the magnetic field lines produced by a coronal shock wave generates the prominence oscillation. In turn, \citet{2021lunaApJ912} by 2.5D MHD modelling demonstrated that coronal jets are able to produce large-amplitude oscillations. Alternatively on an analytical study representing a filament by a current-currying wire, which is similar to the present model, \citet{2016kolotkovA&A590} developed a linear model of the transverse oscillations of a prominence determining periods and stability conditions. This article was followed by a weakly nonlinear and a fully nonlinear study considered in \citet{2018kolotkovJASTP172}. Beyond the differences between the models or the proposed disturbances, there is a consensus that the magnetic force plays an important role as a restoring force characterising the oscillations. However, the analysis of oscillation properties is a topic still open to discussions.

With regards to the damping of oscillations, several theoretical mechanisms were proposed. An important aspect is the interaction between the moving filament and the ambient coronal plasma. This motion gives rise to a process where a transfer of energy from longer scales to smaller ones takes place until it finally dissipates due to viscous effects on smaller scales. Also, viscous processes were described by \citet{1966ZA.....63...78H}, and the damping contribution by emission of magnetoacoustic waves, known as wave leakage, was described by  \citet{1969SoPh....6...72K} and \citet{1992SoPh..142..113V}. For example, \cite{1999A&A...345.1038S}, considering filament oscillations in terms of a solid body, suggested that vertical oscillations lead to fast wave emission carrying momentum away from the filament and so damping the motion, whereas horizontal oscillations would lead to the emission of slow waves but in a less effective process. In this sense, \citet{2018arregieLRSP15} noticed that as emitted slow waves propagate along magnetic field lines are unable to take energy out of an environment with closed magnetic lines. Under these conditions, wave leakage is only possible by fast waves. Another important damping mechanism is the resonant wave process \citep[see  e.g.,][]{2011SSRv..158..289G}, where the energy of the global oscillation is transferred to an inhomogeneous transition layer between the filament and the solar corona. For instance, \cite{2011A&A...531A..53H} observationally studying large-amplitude transverse oscillations of a filament, where periods and damping times for different heights were determined, found that the damping times showed a linear dependence with the periods, where the resonant absorption was suggested as the main damping mechanism. Particularly, throughout 3D numerical simulations \cite{2016ApJ...820..125T} showed evidence that the energy transfer from global oscillations to continuum Alfv\'en modes at the filament's edge is the main damping mechanism.

With the aim to contribute to the understanding of quiescent filament oscillations, in this paper we perform 2.5D ideal MHD simulations. Our motivation is to study large-amplitude transverse oscillations driven by an external coronal shock wave produced by a distant event, e.g., a jet or a flare. For this purpose, we consider different scenarios varying the height, size and mass of the filament. The interaction between the filament and the coronal shock wave is analysed by means of the main motion characteristics, i.e., period of oscillation, amplitudes, damping time and restoring forces. 

The paper is organised as follows: in Section 2 we  present the governing equations, the solar atmospheric scenario and the magnetic model used to represent the filament, detailing how the initial equilibrium conditions were found in a stratified medium with variable gravity. The main features of the numerical code used to run different configurations and the perturbing method are  developed. In Section 3 we  present the results and discussion, starting with the analysis of the filament dynamics in terms of the forces involved to continue with a parametric study evaluating the different configurations. Finally in Section 4 the main conclusions are exposed.

\section{The model}
\label{S-Model}
\begin{figure} 
\centerline{\includegraphics[width=0.8\textwidth]{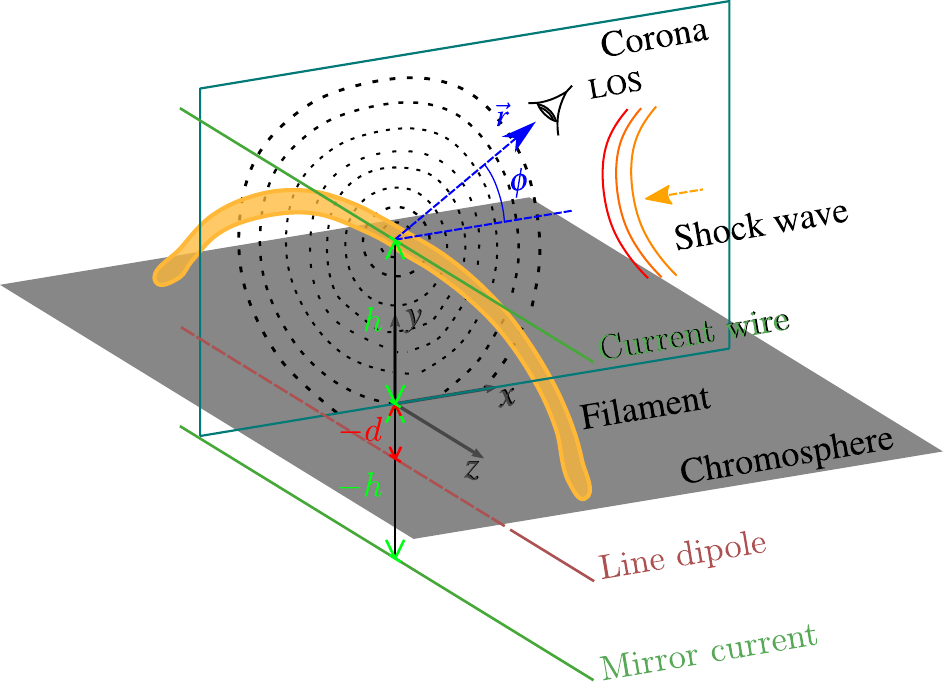}}
\caption{Our schematic scenario of a quiescent filament floating in a stratified atmosphere hit transversely by a coronal shock wave. The plane $(x,y)$ is considered in numerical simulations.} 
\label{f:esquema}
\end{figure}

By numerical experiments we study transverse oscillatory properties of quiescent filaments triggered by a large-scale coronal wave coming from a remote site. The scenario is depicted in Figure~\ref{f:esquema}. The quiescent filament is assumed to be located in a coronal quiet-region, being long enough such that it can be adequately represented by a 2.5D modelling in a perpendicular plane of symmetry. Based on the model of \cite{1990forbesJGR95}, the scenario includes: a filament floating in the corona at a certain height, a variable gravity and a stratified coronal background in hydrostatic equilibrium, with the chromosphere and a thin transition region at the base. The large-scale coronal wave perturbation is emulated by a blast. This scenario allows us to carry on oscillatory analysis measuring periods of oscillations, amplitudes and damping times. 

An ideal MHD scenario is considered with a compressible plasma in presence of a gravitational field. In conservative form the ideal MHD equations read (CGS units):
\begin{eqnarray}
\frac{\partial \rho}{\partial t} + \mathbfit{\nabla}\cdot\left(\rho\mathbfit{v}\right)  &=& 0, \nonumber \\ 
\frac{\partial \rho\mathbfit{v}}{\partial t} + \mathbfit{\nabla}\cdot \left(\rho\mathbfit{v}\mathbfit{v}-\frac{1}{4\pi}\mathbfit{B}\mathbfit{B}\right) &=& -\mathbfit{\nabla}p + \frac{1}{c}\mathbfit{j}\times\mathbfit{B} + \rho\mathbfit{g}, \nonumber \\ 
\frac{\partial\mathbfit{B}}{\partial t} + \mathbfit{\nabla} \cdot \left(\mathbfit{vB}-\mathbfit{Bv}\right) &=& 0, \nonumber \\
\frac{\partial E}{\partial t} + \mathbfit{\nabla}\cdot \left[(E+p+\frac{B^2}{8\pi})\mathbfit{v}-\frac{1}{4\pi}\mathbfit{B}(\mathbfit{v}\cdot\mathbfit{B})\right] &=& \rho\mathbfit{g}\cdot\mathbfit{v}, 
\label{e:mhd}
\end{eqnarray}
where $\rho$ is mass density, $\mathbfit{v}$ is plasma velocity, $\mathbfit{B}$ is magnetic field, $\mathbfit{j}$ is current density and $\mathbfit{g}$ is acceleration of gravity. Also, $E$ is total energy, $e$ is internal energy and $p$ is thermal pressure, whose expressions are 
\begin{eqnarray} 
&& E = \frac{1}{2}\rho v^2 + e + \frac{B^2}{8\pi}, \nonumber \\ 
&& e = \frac{p}{\left(\gamma -1\right)}, \nonumber \\ 
&& p = \frac{R_{\rm{g}}}{\bar{\mu}}\rho T,
\label{e:ener}
\end{eqnarray}
with $T$ the plasma temperature, $\gamma=5/3$ and $R_{\rm{g}}$ the gas constant. The ideal plasma is assumed to be fully ionised and with a solar abundance\footnote{Solar abundance: $70.7\%~$ H $+$ $27.4\%~$ He $+$ $1.9\%~$ heavier elements.} (implying a mean atomic mass $\bar{\mu}=0.613$). The Amp\`ere's law is $\mathbfit{j}=\frac{c}{4\pi}\curl\mathbfit{B}$, with $c$ the speed of light. Furthermore, let us note in the conservative MHD Equations~\ref{e:mhd} appear the tensor products $\mathbfit{v}\mathbfit{v}$, $\mathbfit{vB}$, $\mathbfit{Bv}$ and $\mathbfit{BB}$ (the magnetic part of the Maxwell stress tensor).

\subsubsection{Stratified atmosphere}
\label{sub:sa}
To describe our model set-up two reference systems were used with different origins. First, a Cartesian coordinates $(x,y,z)$ was used to describe the set-up as a whole with its origin fixed at the base of the chromosphere, the $y$-axis pointing radially away from the solar surface and the $x$-, $z$-axes parallel to the surface (neglecting the surface curvature). Second, to locally describe the filament structure it is convenient to use a cylindrical coordinate system $(r,\phi,z')$ with its origin fixed at the filament's centre, for the radial, azimuthal and axial directions, respectively, see Figure~\ref{f:esquema}. Note that as the $z'$- and $z$-axes are parallel, the $z'$ coordinate can be omitted. 

The simulated solar atmosphere is in hydrostatic equilibrium with a gravitational stratification in the vertical direction including a temperature profile for the chromosphere, transition region and corona. The chromosphere has a height $h_{\rm {cr}}=2.5~$Mm with a constant temperature $T_{\rm {cr}}=1.2\times 10^4~$K, the transition region extends up to a height $h_{\rm {tr}}=0.5~$Mm with a linearly varying temperature, and the corona is based at $y=h_{\rm {cr}}+h_{\rm {tr}}$ extending upwards with a constant temperature $T_0=1~$MK. Therefore, the adopted atmospheric temperature is 
\begin{equation}
T(y)=\left\{ 
\begin{array}{ll} 
T_{\rm {cr}}  &  \ {\rm at~} 0\le y \le h_{\rm {cr}}, \vspace{0.2cm}   \\
\frac{\left(T_0-T_{\rm {cr}}\right)}{h_{\rm {tr}}} \left(y-h_{\rm {cr}}\right) + T_{\rm {cr}} & \ {\rm at~} h_{\rm {cr}}< y < h_{\rm {cr}}+h_{\rm {tr}}, \vspace{0.2cm} \\ 
T_0 & \ {\rm at~} y \geq h_{\rm {cr}}+h_{\rm {tr}}. 
\end{array} \right.
\label{e:Tatm}
\end{equation}
The initial pressure due to the hydrostatic equilibrium condition, $\frac{{\rm d}p}{{\rm d}y}=-\rho g$, is
\begin{equation}
p(y)=\left\{ 
\begin{array}{ll} 
p_{\rm {tr}} \exp\left(\frac{\alpha}{T_{\rm {cr}}} \left[ \frac{1}{y+R_{\odot}} - \frac{1}{h_{\rm {cr}}+R_{\odot}} \right] \right) & \ {\rm at~} 0\le y \le h_{\rm {cr}}, \vspace{0.25cm}   \\
p_0\exp\left(-\alpha \int_{h_{\rm {cr}}+h_{\rm {tr}}}^{y} \frac{{\rm d}y'}{T(y')\left(y'+R_{\odot}\right)^2} \right) &  \ {\rm at~} h_{\rm {cr}}< y < h_{\rm {cr}}+h_{\rm {tr}}, \vspace{0.25cm}  \\ 
p_0\exp\left(\frac{\alpha}{T_0} \left[\frac{1}{y+R_{\odot}} - \frac{1}{h_{\rm {cr}}+h_{\rm {tr}}+R_{\odot}}\right] \right) &  \ {\rm at~} y \geq h_{\rm {cr}}+h_{\rm {tr}},
\end{array} \right.
\label{e:Patm}
\end{equation}
with the reference pressure $p_0=\frac{R_{\rm{g}}}{\bar{\mu}}\rho_0 T_0$ fixed at the coronal base and
\begin{equation}
p_{\rm {tr}} = p_0\exp\left(-\alpha \int_{h_{\rm {cr}}+h_{\rm {tr}}}^{h_{\rm {cr}}} \frac{{\rm d}y'}{T(y')\left(y'+R_{\odot}\right)^2} \right) 
\label{e:Ptr}
\end{equation} 
being the extrapolated pressure at the base of the transition region. The gravity acceleration is $\mathbfit{g}(y)=-\frac{G M_{\odot}}{(y+R_{\odot})^2}\hat{\mathbfit{j}}$. $\alpha=\frac{\bar{\mu}GM_{\odot}}{R_{\rm{g}}}$ is a constant, with $G$ the gravitational constant, $M_{\odot}$ the solar mass and $R_{\odot}$ the solar radius. The density $\rho$ and internal energy $e$ in the atmosphere are determined by the Equations of state \ref{e:ener}. Finally, the integral in Equations~\ref{e:Patm} and \ref{e:Ptr} has a closed analytic solution if $T(y)$ is a linear function, which is
\begin{eqnarray}
\int_{y_0}^{y} \frac{{\rm d}y'}{T(y')\left(y'+R_{\odot}\right)^2} &\!=\!&  \frac{1}{R_{\odot}a-b}\left(\frac{1}{y+R_{\odot}}-\frac{1}{y_0+R_{\odot}}\right) 
 \nonumber \\ 
&& 
+ \frac{a}{\left(R_{\odot}a-b\right)^2} \ln\left(\frac{T(y)}{T(y_0)}\frac{y_0+R_{\odot}}{y+R_{\odot}} \right).
\end{eqnarray} 
Constants $a$ and $b$ are obtained by writing the linear temperature of the transition region in Equation \ref{e:Tatm} as $T(y)=ay+b$, having $a\equiv \frac{T_0-T_{\rm {cr}}}{h_{\rm {tr}}}$ and $b\equiv T_{\rm {cr}}-a h_{\rm {cr}}$.

\subsubsection{The filament} 
\label{sub:fil}
In line with \citet{1990forbesJGR95}, but adding the gravity force, the flux rope is modelled as an electric current-carrying wire of radius $R$ floating in the corona. The magnetic configuration consists of three components: the current-carrying wire located at height $y=h$; a mirror current at depth $y=-h$; and a line dipole of relative strength $M_{\rm dip}$ at depth $y=-d$, representing the photospheric field. 
The initial magnetic field, considering each contribution respectively, is:
\begin{eqnarray}
B_x(x,y) &=& - \frac{(y - h)}{r_-}B_\phi(r_-) + \frac{(y + h)}{r_+}B_\phi(r_+) \nonumber \\
         &&   + \frac{\left(x^2 - (y + d)^2\right)}{r_d^4} M_{\rm{dip}} d R_3 B_\phi (R_3) , \nonumber \\
B_y(x,y) &=& \frac{x}{r_-}B_\phi(r_-) - \frac{x}{r_+}B_\phi(r_+) + \frac{2x(y + d)}{r_d^4} M_{\rm{dip}} d R_3  B_\phi (R_3), \nonumber  \\ 
B_z(x,y) &=& B_z(r),
\label{e:bxy}
\end{eqnarray}
with $r_{-}$, $r_{+}$ and $r_d$ being the distances from the filament, the mirror current and the dipole, respectively, to the observed point, where
\begin{eqnarray}
r_\pm &=& \sqrt{x^2 + (y \pm h)^2},  \\
r_d   &=& \sqrt{x^2 + (y + d)^2}. 
\label{e:rB}
\end{eqnarray}
$B_\phi$, $B_z$, $R_3$ and $M_{\rm dip}$ are defined next. 

We assume a transition layer of thickness $\Delta$ to relate the internal structure of a filament of radius $R$ with the outer coronal region. Thus, three filament zones are distinguished: an inner zone (Z1) for radius $0\le r \le R_2$; a transition zone (Z2) for radius $R_2 < r < R_3$; and the external zone (Z3) for $r\ge R_3$. $R_2=R-\frac{\Delta}{2}$ and $R_3=R+\frac{\Delta}{2}$. High-resolution observations have revealed the existence of inhomogeneous fine structures of filaments \citep{2005SoPh..226..239L}, making reasonable to include in the model a non-uniform layer between the filament inner zone and the coronal external zone \citep[e.g., as done by][and references therein]{2009solerApJL}.

The assumed electric current density inside the filament is $\mathbfit{j}(r)=j_\phi${\boldmath$\hat{\phi}$}$+ j_z  \hat{\mathbfit{k}}$, with  components: 
\begin{equation}
\! j_z(r) \!=\! \left\{ 
\begin{array}{ll} 
j_0 & \vspace{0.2cm} \\
\! \frac{j_0}{2} \left(\cos\left[\frac{\pi}{\Delta}\left(r-R_2\right)\right] + 1 \right) & \vspace{0.2cm} \\ 
0  &
\end{array} \right.
{\rm {and}} \quad 
j_\phi(r) \!=\! \left\{ 
\begin{array}{ll} 
\! \frac{j_1 r}{2\sqrt{R_2^2-r^2}},  & \ : \ {\rm at~Z1}, \vspace{0.2cm} \\ 
\! 0 & \ : \ {\rm at~Z2}, \vspace{0.2cm}  \\
0  & \ : \ {\rm at~Z3},
\end{array} \right.
\label{e:jz}
\end{equation}
where $j_0$ and $j_1$ are constants. The azimuthal component of the magnetic field $B_\phi(r)$ produced by the axial current density $j_z$ is
\begin{equation}
B_\phi(r)=\left\{ 
\begin{array}{ll} 
\frac{2\pi j_0}{c} r & \ {\rm at~Z1}, \vspace{0.2cm}  \\
\frac{2\pi j_0}{c}\frac{1}{r} \left\lbrace \frac{r^2}{2} + \frac{R_2^2}{2} - \left(\frac{\Delta}{\pi}\right)^2 +  \left(\frac{\Delta}{\pi}\right)^2 \cos\left[\frac{\pi}{\Delta}\left(r-R_2\right)\right] \right.   \\ 
\left. +\frac{\Delta}{\pi}r \sin\left[\frac{\pi}{\Delta}(r-R_2)\right] \right\rbrace  & \ {\rm at~Z2}, \vspace{0.2cm} \\ 
\frac{2\pi j_0}{c}\frac{1}{r} \left(\frac{R_3^2}{2} + \frac{R_2^2}{2} - 2\left(\frac{\Delta}{\pi}\right)^2\right) & \ {\rm at~Z3}, 
\end{array} \right.
\label{e:bphi}
\end{equation}
and the axial component is
\begin{equation}
B_z(r)=\left\{ 
\begin{array}{ll} 
\frac{\sqrt{8}\pi}{c}j_1\sqrt{R_2^2-r^2}  &  \ {\rm at~Z1}, \vspace{0.2cm}  \\
0 &  \ {\rm at~Z2}, \vspace{0.2cm} \\ 
0 &  \ {\rm at~Z3}.
\end{array} \right.
\label{e:bz}
\end{equation}

The initial thermal pressure inside the filament, intending to be in equilibrium as close as possible   with its neighbourhood and satisfying $\mathbfit{\nabla}p - \frac{1}{c}\mathbfit{j}\times\mathbfit{B}\approx 0$, results in
\begin{equation} 
 p(r) = p_{\rm c}(r) - \frac{1}{c}\int_r^\infty j_\phi(r')B_z(r') \, {\rm d} r' + \frac{1}{c}\int_r^\infty j_z(r')B_\phi(r') \, {\rm d}r',
\label{e:p1}
\end{equation}
where $p_{\rm c}$ is the background coronal pressure given by the last expression of Equation \ref{e:Patm}. Considering the different shells in the radial direction throughout the filament, the pressure is 
\begin{equation}
p(r)=\left\{ 
\begin{array}{ll}
p_{\rm c}(r) + \frac{\pi}{c^2}\left(j_0^2-j_1^2\right)\left(R_2^2-r^2\right) 
+ \frac{1}{c}\int_{R_2}^{R_3} j_z(r')B_\phi(r') \, {\rm d}r'  &  \ {\rm at~Z1}, \vspace{0.2cm} \\
p_{\rm c}(r) + \frac{1}{c}\int_{r}^{R_3} j_z(r')B_\phi(r') \, {\rm d}r' & \ {\rm at~Z2},  \vspace{0.2cm} \\ 
p_{\rm c}(r) & \ {\rm at~Z3}.
\end{array} \right.
\label{e:Pfil}
\end{equation}
This balance pressure was obtained taking into account the azimuthal magnetic field $B_{\phi}$ generated by the filament, written in Equation~\ref{e:bphi}, but neglecting the external contributions to the azimuthal component by the mirror wire and the dipole. The prescribed filament temperature is 
\begin{equation}
T(r)=\left\{ 
\begin{array}{ll} 
T_{\rm {fil}}  &  \ {\rm at~Z1}, \vspace{0.2cm}  \\
\frac{\left(T_0-T_{\rm {fil}}\right)}{\Delta} \left(r-R_2\right) + T_{\rm {fil}} &  \ {\rm at~Z2}, \vspace{0.2cm}  \\ 
T_0 &  \ {\rm at~Z3},
\end{array} \right.
\label{e:Tfil}
\end{equation}
with the constant inner temperature of the filament $T_{\rm {fil}}=0.3~$MK. Finally, the density $\rho(r)$ of the filament is obtained introducing the expressions for the pressure and temperature, Equations \ref{e:Pfil} and \ref{e:Tfil}, into Equation \ref{e:ener}. 

\begin{figure}    
\centerline{
\hspace*{0.015\textwidth}
\includegraphics[width=0.59\textwidth]{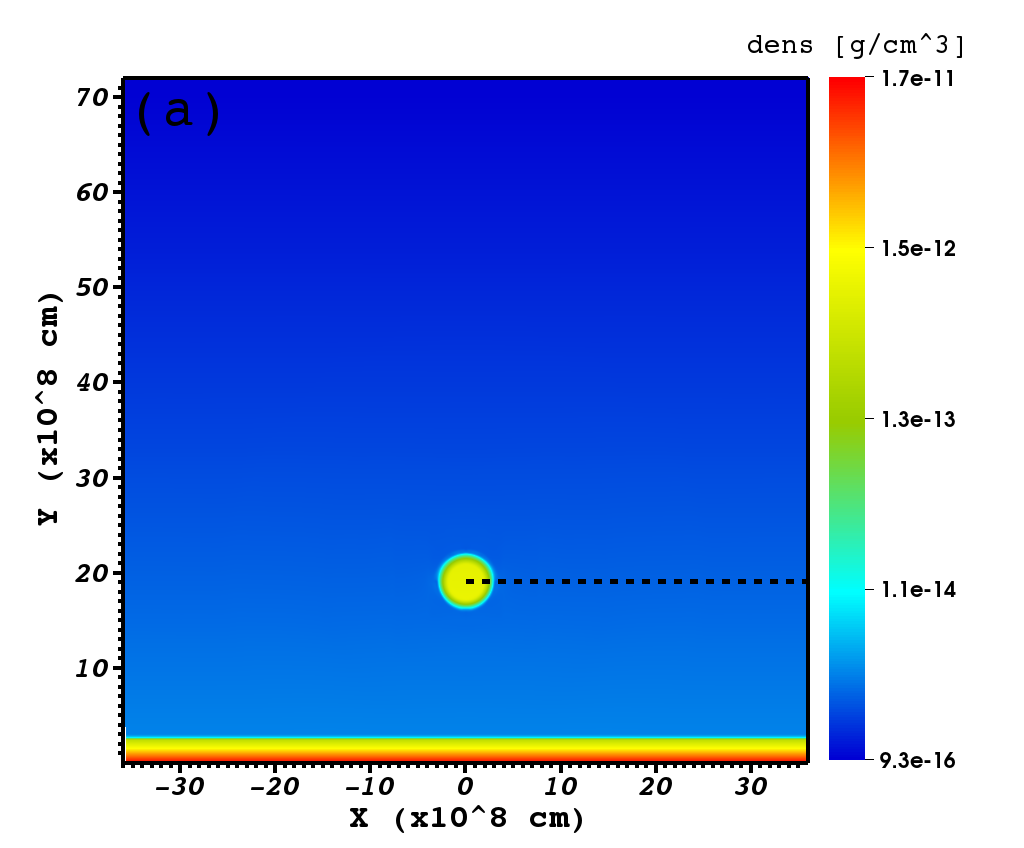}
\hspace*{-0.05\textwidth} 
\includegraphics[width=0.58\textwidth]{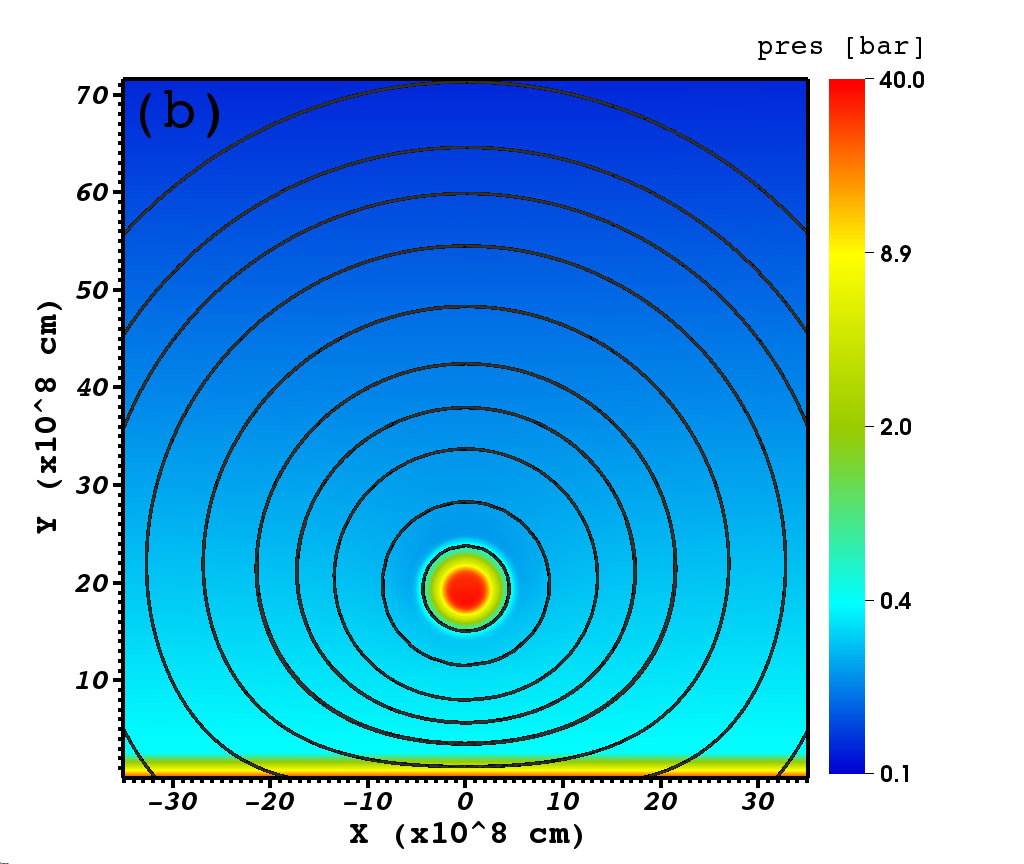}}
\caption{Zoom-in snapshots of the initial conditions for the filament floating in a stratified atmosphere. \textsf{(a)} The density pattern is shown. \textsf{(b)} Thermal pressure with total magnetic field lines superimposed (black-solid lines). The thick-dashed line in panel (a) indicates the slice used to evaluate the distance-magnetic field plot. This filament configuration corresponds to Case 3 defined below in Section \ref{Sub:sim}.}
\label{f:init_cond}
\end{figure}

Figure~\ref{f:init_cond} shows zoom-in initial conditions  for a particular-filament case floating in the stratified atmosphere. The density pattern is shown in panel \textsf{(a)}, and the thermal pressure with the total magnetic field lines superimposed is shown in panel \textsf{(b)}.

\subsubsection{Equilibrium condition} 
\label{sub:equi}
In order to establish equilibrium conditions, the forces acting on the filament are analysed. The forces are: the upward force exerted by the mirror current $F_{\rm mir}$; and the downward forces of the dipole $F_{\rm dip}$ and the weight $F _{\rm g}$. Thus, the initial total force $F$, per unit length $L$,  in the vertical direction is
\begin{eqnarray}
\frac{F}{L} &=& F_{\rm mir} - F _{\rm dip} - F_{\rm g}  \nonumber \\ 
            &\approx& \frac{1}{c^2}\frac{I^2}{h} - \frac{2}{c^2}\frac{Im}{\left(h+d\right)^2} - \left(m_{\rm fil} - m_{\rm buo}\right)g, 
            \label{e:force}
\end{eqnarray}
where approximations were made on the magnetic force expressions and are valid for a small radius $R$. $I$ is the electric current flowing inside the filament and its mirror, $m$ is the  dipole strength, $m_{\rm fil}$ is the mass of the filament (per unit length) and $m_{\rm buo}$ is the background mass enclosed in a filament-like volume. The weight term  includes the buoyancy force $m_{\rm buo}g$ that is smaller than the weight, $\sim$1$\%$. The axial electric current going through the transverse filament area $A_{\rm fil}$ is
\begin{equation}
I=\int_{A_{\rm fil}} j_z \hat{\mathbfit{k}}\cdot{\rm d}\mathbfit{A'}= \frac{\pi j_0}{2}\left(R_3^2+R_2^2-\left(\frac{2\Delta}{\pi}\right)^2 \right).
\label{e:curr}
\end{equation}
%

\begin{figure} 
\centerline{\includegraphics[width=0.6\textwidth]{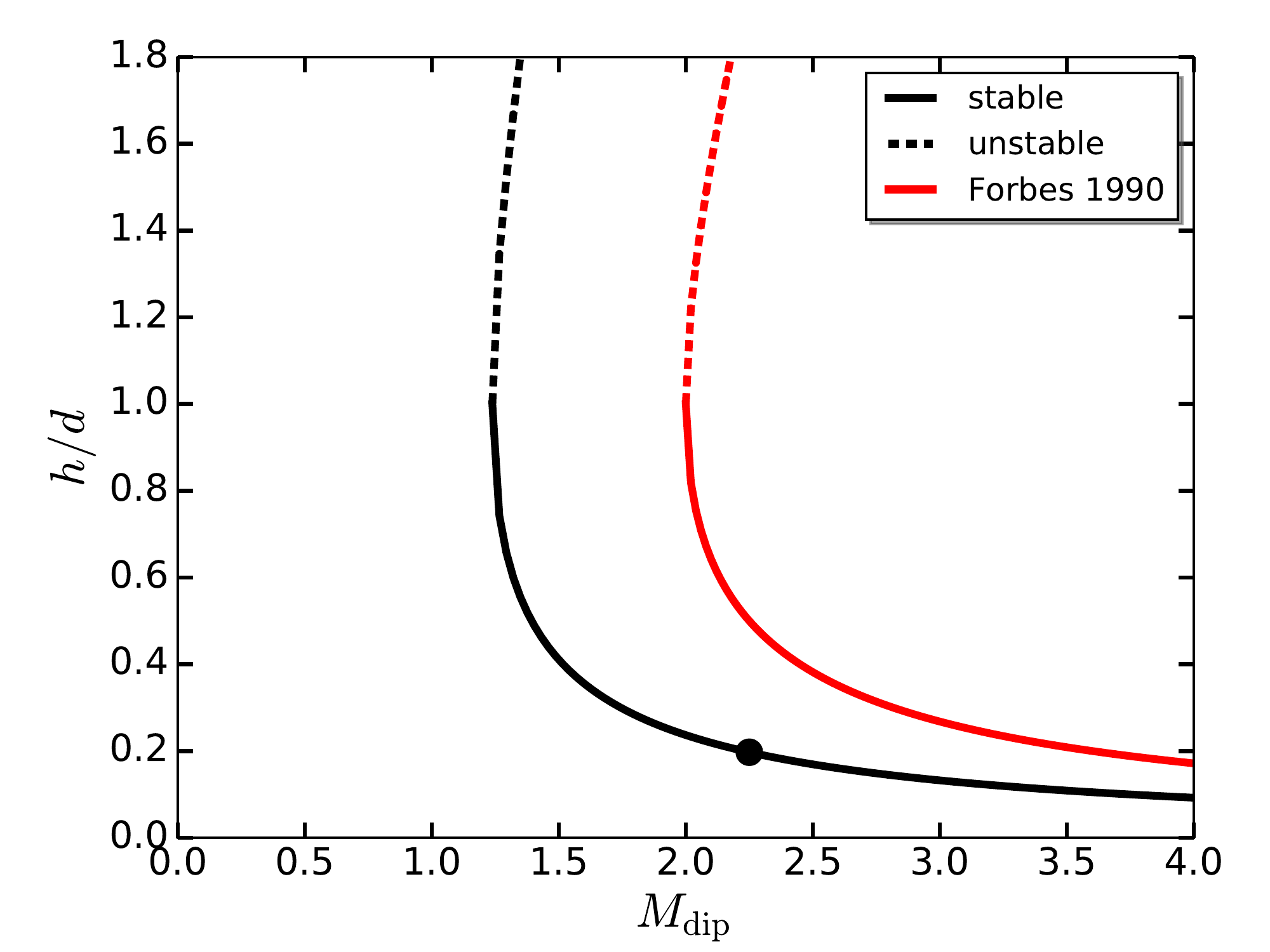}}
\caption{The locations of equilibria for the ratio $\frac{h}{d}$ as a function of the relative strength of the dipole $M_{\rm{dip}}$. Black-solid portion of the line represents stable equilibria, while black-dashed portion unstable ones. Red lines display the same example but neglecting the gravity. The equilibria indicated by the black lines and the stable point chosen at $M_{\rm{dip}}=2.25$ correspond specifically to Case 3 defined in Section \ref{Sub:sim}.}
\label{f:equil}
\end{figure}

To satisfy the balance ($F=0$) the following relation must hold:
\begin{equation}
\frac{h}{d}=\frac{\left(M_{\rm{dip}}-(1-M_{\rm{g}})\right)\pm \sqrt{M_{\rm{dip}}^2-2M_{\rm{dip}}(1-M_{\rm{g}})}}{1-M_{\rm{g}}},
\label{e:balan}
\end{equation}
which gives the equilibria in turns of the height and depth as a function of two parameters, $M_{\rm{dip}}$ and $M_{\rm{g}}$. The dimensionless parameter $M_{\rm{dip}}=\frac{m}{Id}$ gives the relative strength between the dipole force $F_{\rm dip}$ and the mirror force $F_{\rm mir}$. Similarly, $M_{\rm{g}}=\frac{c^2 h}{I^2}F_{\rm g}$ is the relative strength between the weight and the mirror forces. Solutions with minus (plus) sign represent stable (unstable) equilibria \citep{1990forbesJGR95}. An example of these equilibria are shown in Figure~\ref{f:equil}, where the radio $\frac{h}{d}$ between the filament height and the dipole depth is plotted as a function of $M_{\rm{dip}}$. The black-solid portion of the line represents stable locations while the black-dashed portion unstable ones. To have an idea of the weight contribution, the red curve represents the equilibria neglecting the weight force. Note that the weight force allows to consider smaller values of the relative dipole strength $M_{\rm{dip}}$, i.e. considering  gravity, the minimum stable location occurs for $M_{\rm{dip}}=1.24$, while neglecting gravity, the minimum stable location occurs for $M_{\rm{dip}}=2$. 

When searching for equilibrium configurations the mass of the filament plays an important role, but its value $m_{\rm fil}=\int_{V_{\rm fil}} \rho(\mathbfit{r'}) \, {\rm d}V'$ has not a closed expression. Therefore, we obtained the following approximated expression
\begin{equation}
m_{fil} \approx \frac{1}{T_{\rm fil}} \left[\pi\rho_0 T_0 R^2 + \epsilon R^4(j_0^2 - \frac{j_1^2}{2}) \right],
\label{e:mfil}
\end{equation}
which is helpful to understand the mass dependence on the set-up parameters. With $\epsilon=\frac{\pi\bar{\mu}}{2R_{\rm g}c^2}$ being a constant. The filament mass increases for larger values of the radius $R$ and the axial current density $j_0$, while decreases with $j_1$. In addition, the mass is inversely proportional to its temperature $T_{\rm fil}$, resulting in heavier filaments for colder temperatures (an vice-versa). In a  lesser degree the mass also depends on the coronal background density $\rho_0$ and temperature $T_0$.

\subsubsection{Perturbation}
\label{sub:pert}
After a relaxation period, the filament was perturbed using a blast mechanism \citep{2004balsaraApJS151}, placed far in the domain's border. This configuration emulates a standing quiescent filament perturbed by an external large-scale coronal wave coming, for example, from a remote energetic flaring site. More details of the simulations are given below in Section \ref{Sub:code}. The wavefront, with a small inclination angle from the horizontal axis towards the solar surface, hits the filament exciting simultaneously horizontal and vertical transverse oscillations. The downward inclination of the wave is in line with observations \citep{2013ApJ...773..166L}. An inclination angle of $\sim$7$^{\circ}$ was used in all cases. The instantaneously applied blast was imposed at $t=30.08~$min, placed at $(x,y)=(95,h+6)~$Mm with a diameter of $20~$Mm and releasing an energy of $7\times 10^{19}~$erg. The average propagation speed of the shock wave was of $486~$km s$^{-1}$. Which in terms of the Alfv\'en Mach number $M_{\rm A}$, for the different cases carried on (see Section \ref{Sub:sim} below) involving different magnetic field strengths (Figure~\ref{f:B_cut}), represented $M_{\rm A}=1.8$ for filaments with small radius and $M_{\rm A}=0.85$ for big radius filaments (in both cases measured at coordinates $(x,y)=(49,19)~$Mm where the ambient magnetic field magnitude behind the wave-front was $3.~$G on the former case and $5.8~$G on the latter).

\cite{2018zhouApJ856} discarded simulations of shock waves as  external perturbations on the filament, by assuming that perturbing the velocity field would be representative of anything that results in a bulk motion of the filament. Following this idea, as an alternative to the blast mechanism, we also tried perturbing the velocity field inside and around the filament, which is a much less numerically demanding mechanism. This mechanism is supposed to mimic the velocity impulsive phase exerted by the blast on the filament. However, we found that the resulting oscillations were not equivalent to those obtained from the blast, being the main difference that the damping is stronger when the velocity mechanism was used. Hence, to model large-scale coronal wave perturbations, we discarded the velocity mechanism.

\subsection{Numerical simulations} 

\subsubsection{Code}
\label{Sub:code}
To carry out 2.5D ideal MHD numerical simulations the set of Equations~\ref{e:mhd} was solved using the {\sc Flash} code  \citep[][release 4.5]{2013DubeyIJHPCA28}. This code uses Godunov-type schemes in a co-located regular grid of finite volumes to solve the coupled compressible MHD equations with adaptive mesh refinement (AMR) and high performance computing capabilities. 

For our simulations we chose the unsplit staggered mesh solver with a second-order MUSCL-type reconstruction. This solver, which is based on the scheme by \citet{2009leeJCoPh228}, uses a directionally unsplit technique for the evaluation of numerical fluxes and implements the constrained transport method and the corner transport upwind method for the treatment of the magnetic field, which leads to a better numerical behaviour of the divergence free condition ($\mathbfit{\nabla}\cdot\mathbfit{B}=0$). The Riemann's problems at interfaces of computational cells were calculated using the Roe's solver and the MC slope-limiter was used for reconstructions of cell-centred variables.

The 2D physical domain was delimited by $[-120, 120]~\rm{Mm}\times [0, 120]~\rm{Mm}$ and discretised by a Cartesian grid with eight levels of refinement with $20 \times 10$ cells per block, which gives a maximal spatial resolution of $\delta x = \delta y = 0.094~$Mm. Boundary conditions were set as follows: for the left and right borders were chosen zero-gradient (outflow) conditions for thermodynamic variables and velocity, allowing waves to leave the domain without reflection. For the magnetic field at lateral and upper ends we extrapolated the initial condition to ghost cells in order to avoid spurious magnetic forces generated by the violation of the force-free condition of the border background magnetic field. Obviously, this implementation is valid provided that the magnetic perturbations do not reach these borders, although relatively small numerically induced forces can be tolerated since the region of interest is at  the  prominence location. The bottom and upper borders require more detailed treatment due to the presence of gravity and the tying of the magnetic field lines to the chromosphere. For the vertical extrapolation of thermodynamic variables we implemented the hydrostatic boundary conditions proposed by \citet{2019krauseAA631} with constant temperature, while the zero-gradient condition is imposed for the velocity at the upper end. On the other hand, the line-tied boundary conditions proposed by \citet{1987SoPh..114..311R} were applied to the magnetic field and velocity at the bottom. 

Taking into account that our study requires an accurate model of the prominence equilibrium at the unperturbed condition, we needed to improve the capability of the numerical code to preserve the hydrostatic equilibrium. As explained by \citet{2019krauseAA631}, the traditional MUSCL-type scheme are not able to satisfy that condition since the vertical exponential decay of the pressure cannot be exactly approximated by a polynomial reconstruction. In addition, the numerical pressure gradient is not cancelled by the gravitational source term when a simple cell-centred evaluation is used. Consequently, spurious momentum fluxes are numerically induced producing non-physical velocities. To avoid this undesirable behaviour, we implemented the local hydrostatic reconstruction scheme and we improved the equilibrium preservation during the simulation.

\subsubsection{Simulations}
\label{Sub:sim}
The overall simulations were composed by two stages. The first one was the relaxation period beginning from the initial condition at $t_0 = 0$ (see Figure~\ref{f:init_cond}) and extending for 30 min to allow an equilibrium adjustment between the filament and its background environment. This time lapse is necessary because  initially the filament is not in exact equilibrium. After this  time lapse the system reaches a pseudo-equilibrium position, stationary enough to be considered still. By the end of this first period the non-equilibrium velocity of the filament is smaller than $1~$km s$^{-1}$, that means an order of magnitude smaller than the velocity exerted by the blast. The second stage corresponds to the oscillating scenario, which starts when  the blast is discharged, at $t_{\rm pert}=30.08~$min, and lasts until the simulation ends. During this stage the shock wave generated by the blast passed through the filament and the oscillating motion is triggered.

Aiming to make a seismological study of the system as a function of height, size and mass of the filament, seven simulations were performed, six of them for three different heights ($h=[14,19,25]~$Mm) and two radii ($R=[2.5,3.5]~$Mm), and the remaining one varying the mass by means of the current density $j_1$. These three parameters were assumed as the independent ones, while the remaining parameters, i.e. $j_0$, $\rho_0$, $\Delta$, $M_{\rm dip}$, $T_0$, $T_{\rm cr}$, $T_{\rm fil}$, $h_{\rm cr}$ and $h_{\rm tr}$, have fixed values for all cases. The  details of each case and the fixed parameters  are listed in Table~\ref{t:fits} (Set-up parameters). For all cases: $j_0 = 600~$statA cm$^{-2}$; $\rho_0 = 3.2\times 10^{-15}~$g cm$^{-3}$; $\Delta = 1.25~$Mm; and $M_{\rm{dip}} = 2.25$. Note that the azimuthal current density $j_1=600~$statA cm$^{-2}$ is the same for all cases, except in Case 5 that is equal to $j_1=797.5$ statA cm$^{-2}$. It should be taken into account that the equilibrium black lines displayed in Figure~\ref{f:equil} are different for all cases. 

\begin{figure}
\centerline{\includegraphics[width=0.6\textwidth]{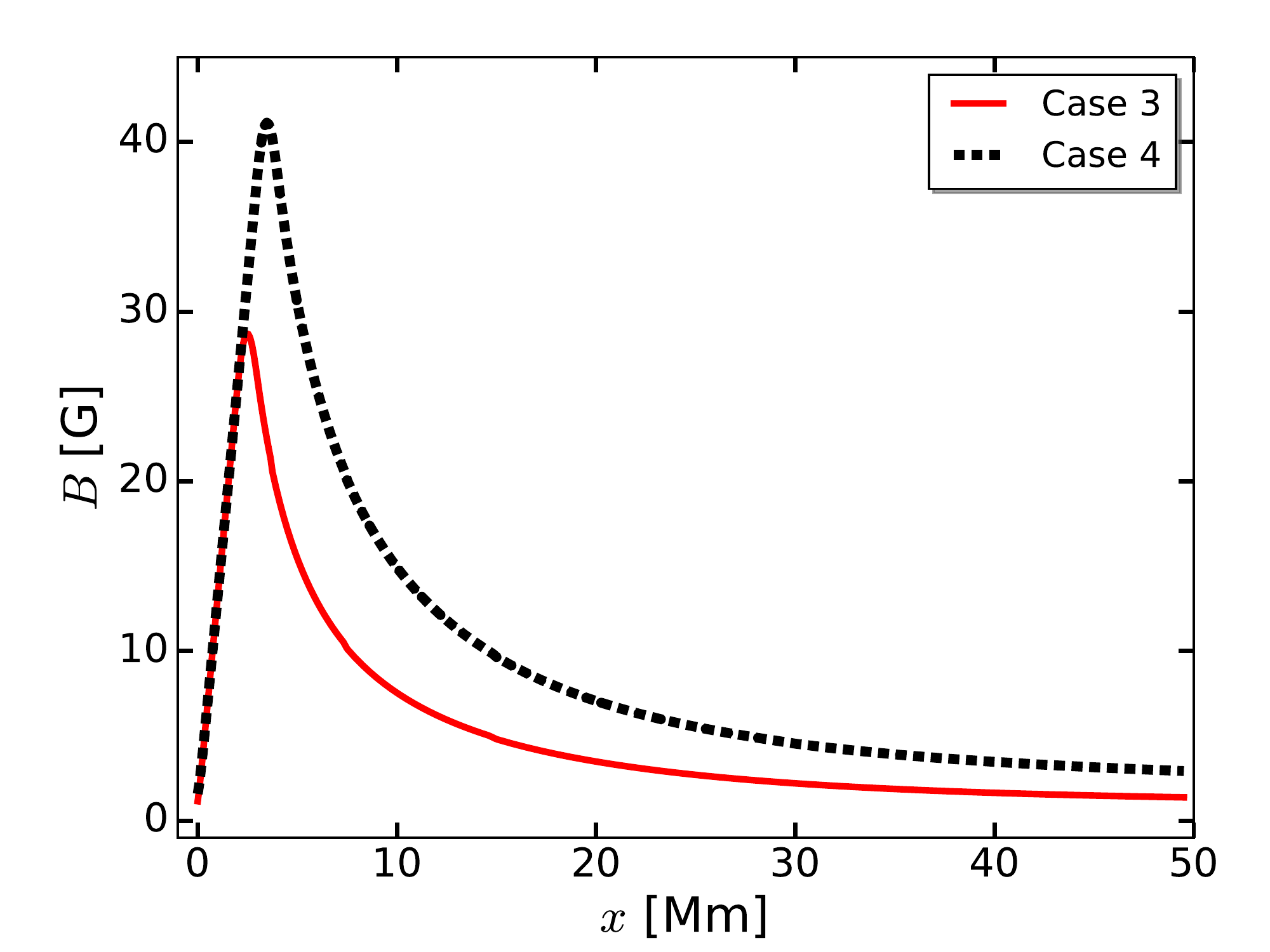}}
\caption{Distance-magnetic field plot along the thick-dashed line of Figure~\ref{f:init_cond}\textsf{(a)}. The values corresponds to the magnetic field magnitude in the plane $(x,y)$ for the filaments of Case 3 (black-solid line) with small radius $R=2.5~$Mm and Case 4 (red-solid line) with big radius $R=3.5~$Mm.}
\label{f:B_cut}
\end{figure}

For all cases, the dipole depth $d$ and strength $m$ are dependent parameters determined by the balance Equation \ref{e:balan}. Thus, to set up equilibrium configurations, higher filaments correspond to deeper magnetic dipoles but interestingly with larger strengths $m$. On the other hand, to increase the filament radius $R$ implies larger values of the magnetic field in the filament surroundings (as displayed in Figure~\ref{f:B_cut}) and larger repelling mirror magnetic forces $F_{\rm mir}$. In addition, a larger radius produces a shallower dipole with larger strength $m$, and also increases the filament mass $m_{\rm fil}$, thus balancing the repelling mirror force. Finally, given the total pressure balance between the filament's interior and exterior by Equation \ref{e:Pfil}, a rise in the azimuthal current density $j_1$, as in Case 5, results in a decrease of the filament mass, generating a shallower dipole with smaller strength $m$.

\begin{figure}    
\centerline{
\hspace*{0.015\textwidth}
\includegraphics[width=0.65\textwidth,angle=0]{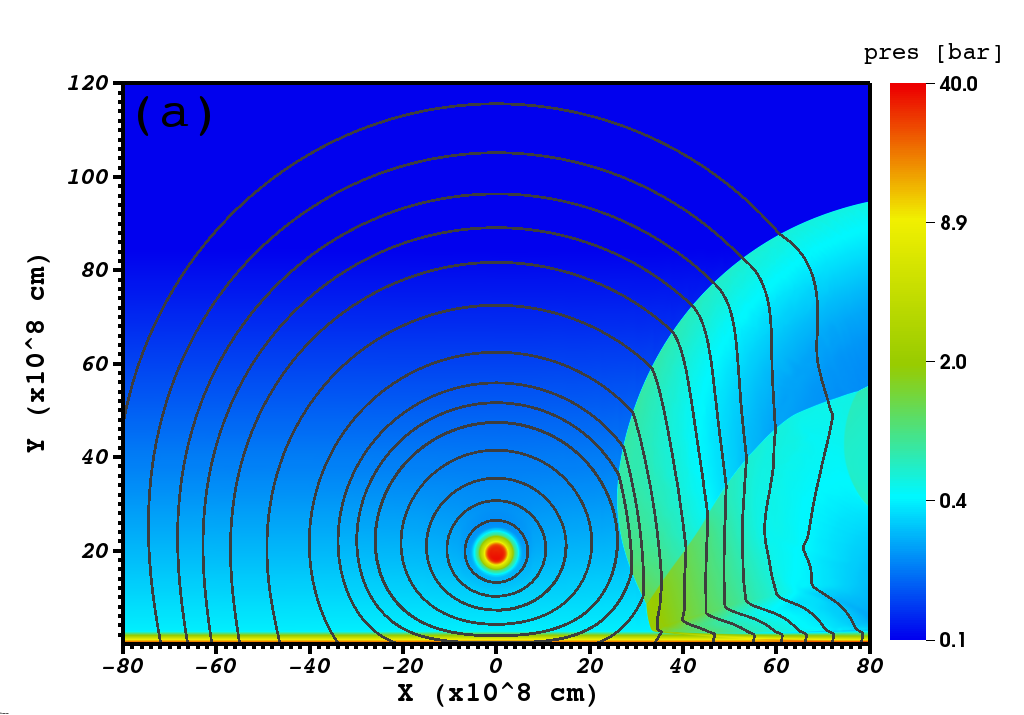}
\hspace*{-0.034\textwidth}
\includegraphics[width=0.65\textwidth,angle=0]{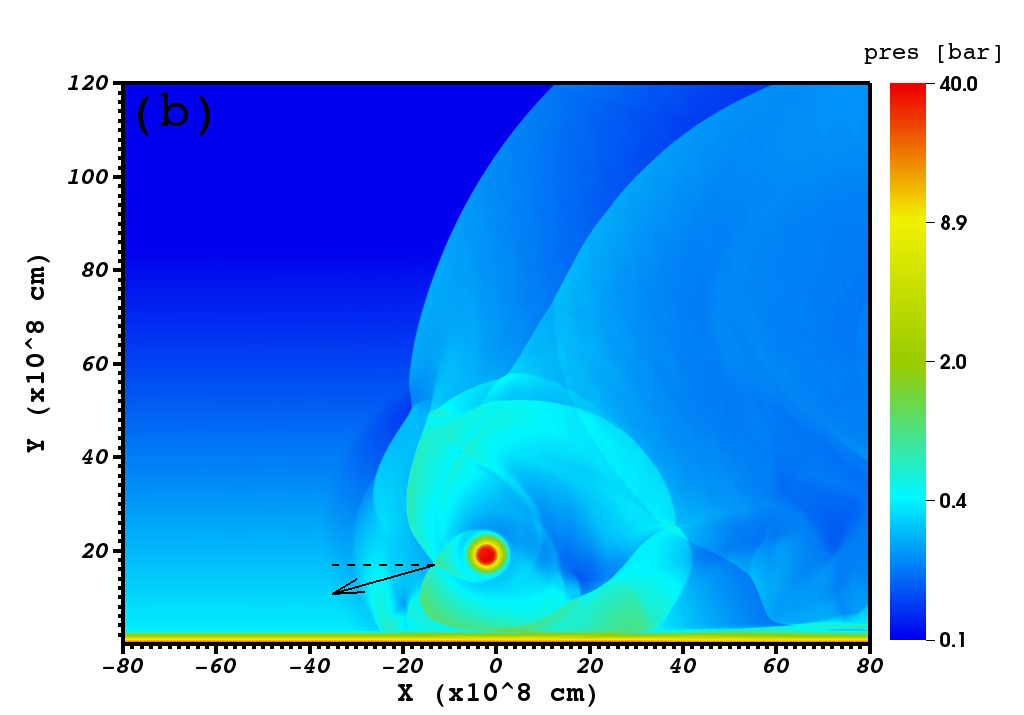}}
\centerline{
\hspace*{0.015\textwidth}
\includegraphics[width=0.65\textwidth,angle=0]{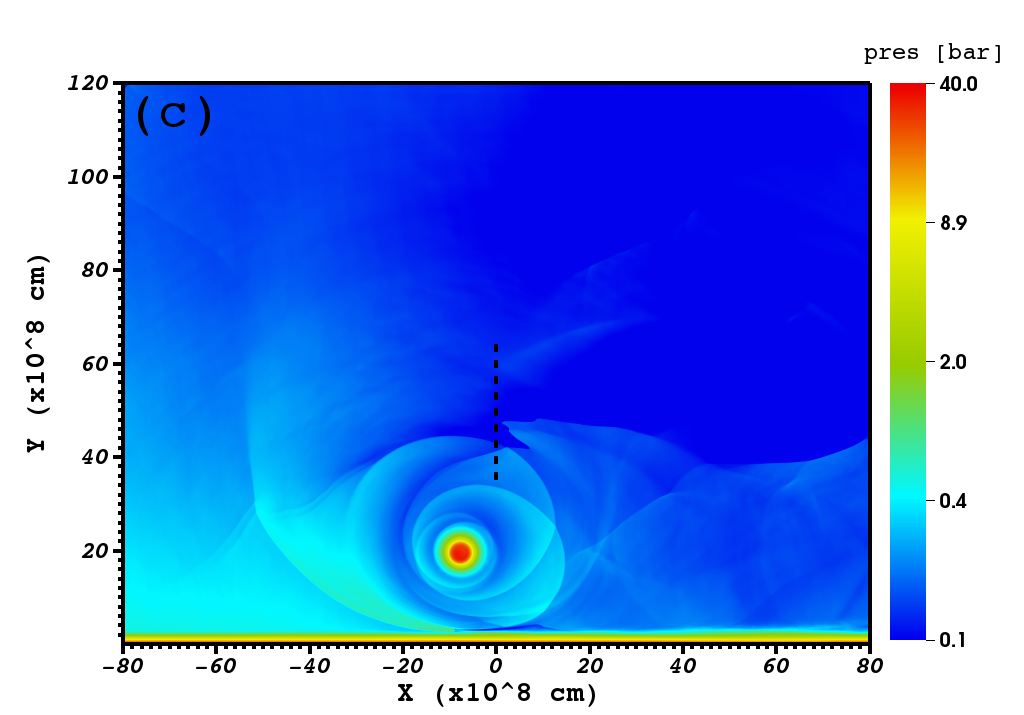}
\hspace*{-0.034\textwidth}
\includegraphics[width=0.65\textwidth,angle=0]{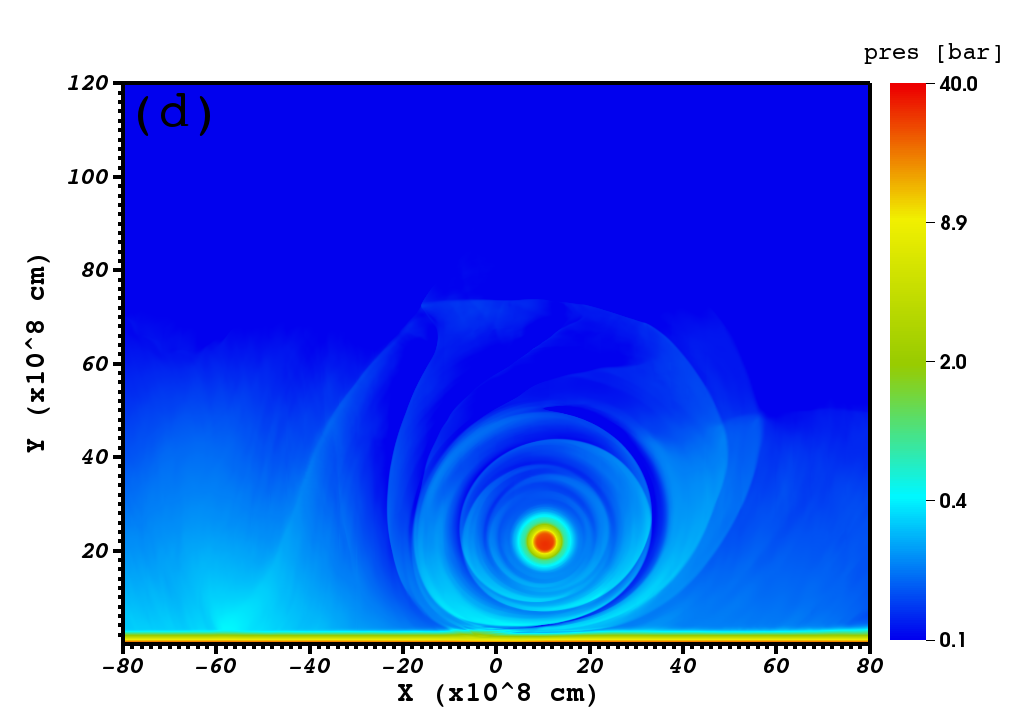}}
\caption{Temporal evolution of the interaction between the quiescent filament and a large-scale coronal wave coming from a remote site. The images correspond to zoom-in snapshots of Case 3. \textsf{(a)} In colour map it is plotted the pressure pattern at time $t=32~$min before the filament-shock interaction begins. The black lines in (a) display the field lines of the total magnetic field. \textsf{(b)} It is shown just a moment after the interaction begins at $t=35~$min. The arrow points the inclination angle of the shock and the dash line represents the horizontal direction. \textsf{(c)} When the filament oscillation reaches its maximum left horizontal displacement at $t=41~$min. Here the vertical dashed line indicates the slice used to evaluate the pressure stack Figure~\ref{f:damping}\textsf{(b)}. \textsf{(d)} Same as (c) but for the maximum right horizontal displacement at $t=59~$min. (See Case 3's movie attached).}
\label{f:pres_patt}
\end{figure}

\section{Results and Discussions} 
\subsection{Dynamics} 
As a result of the interaction with the shock wave, the filament exhibits a damped oscillatory motion coupled in the horizontal and vertical directions. Figure~\ref{f:pres_patt} displays the evolution of the wavefront--filament interaction (see the attached Case 3's movie). Panel \textsf{(a)} shows the shock wave triggered by the blast, travelling at a coronal level and disturbing the chromosphere. In the figure, black lines represent the total magnetic field and the colour map the thermal pressure. On the right side, the presence of waves compressing the plasma and deforming the magnetic field is noticeable. At $t=33~$min, both the main coronal wavefront and its subsequent chromospheric reflection impact the filament and destabilise its equilibrium state (see panel \textsf{(b)}). The main wavefront hits the filament with a small downward inclination angle. Then, the shock wave continues propagating beyond the filament position, dragging it towards the bottom-left direction pointed by the arrow on panel \textsf{(b)}. Later on, the filament starts oscillating during a few cycles (panels \textsf{(c)}--\textsf{(d)}). By $t= 42.5~$min the external perturbation has left the domain. 

\begin{figure}    
\centerline{
\hspace*{0.015\textwidth}
\includegraphics[width=0.515\textwidth]{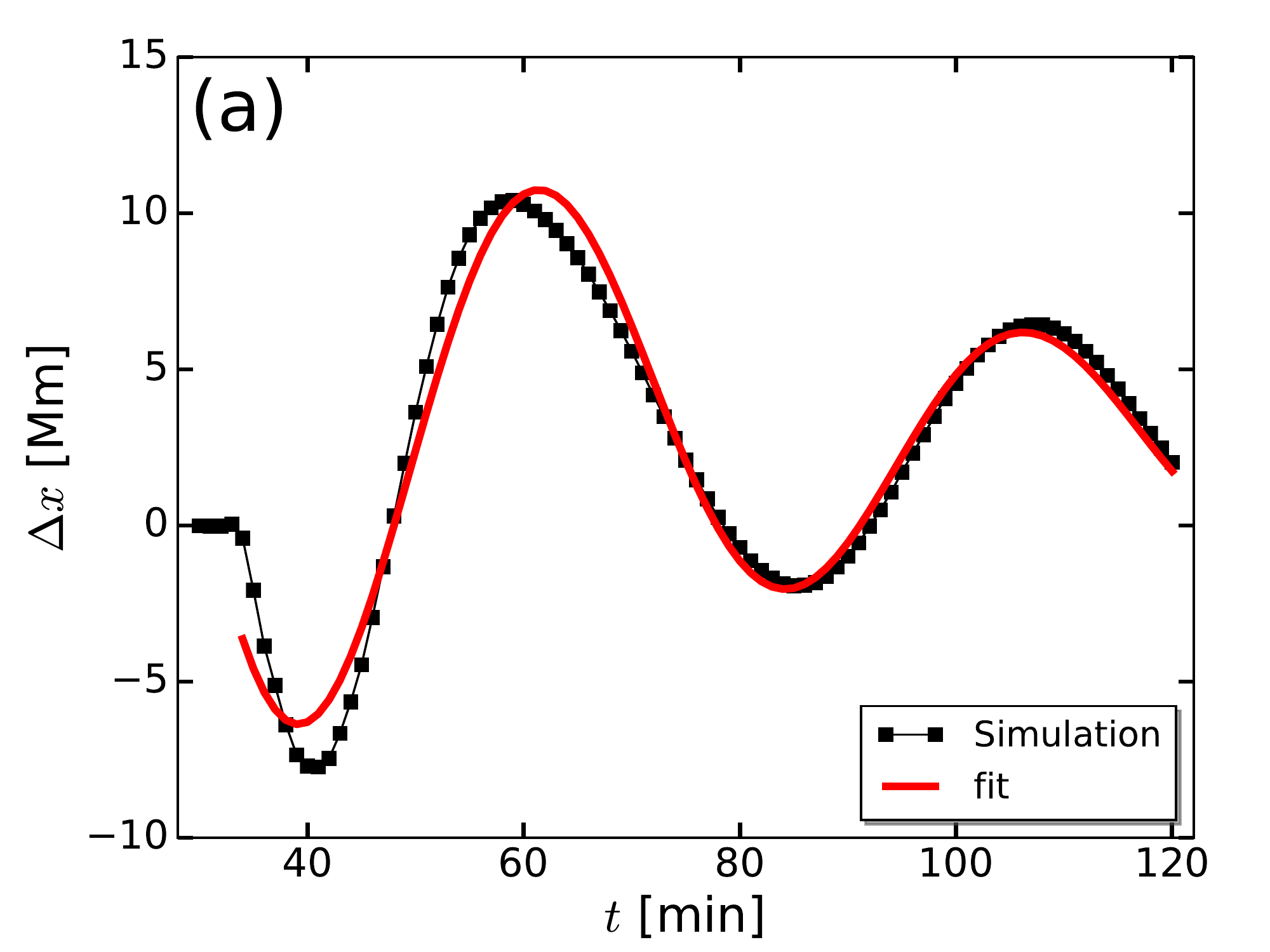}
\hspace*{-0.03\textwidth}
\includegraphics[width=0.515\textwidth]{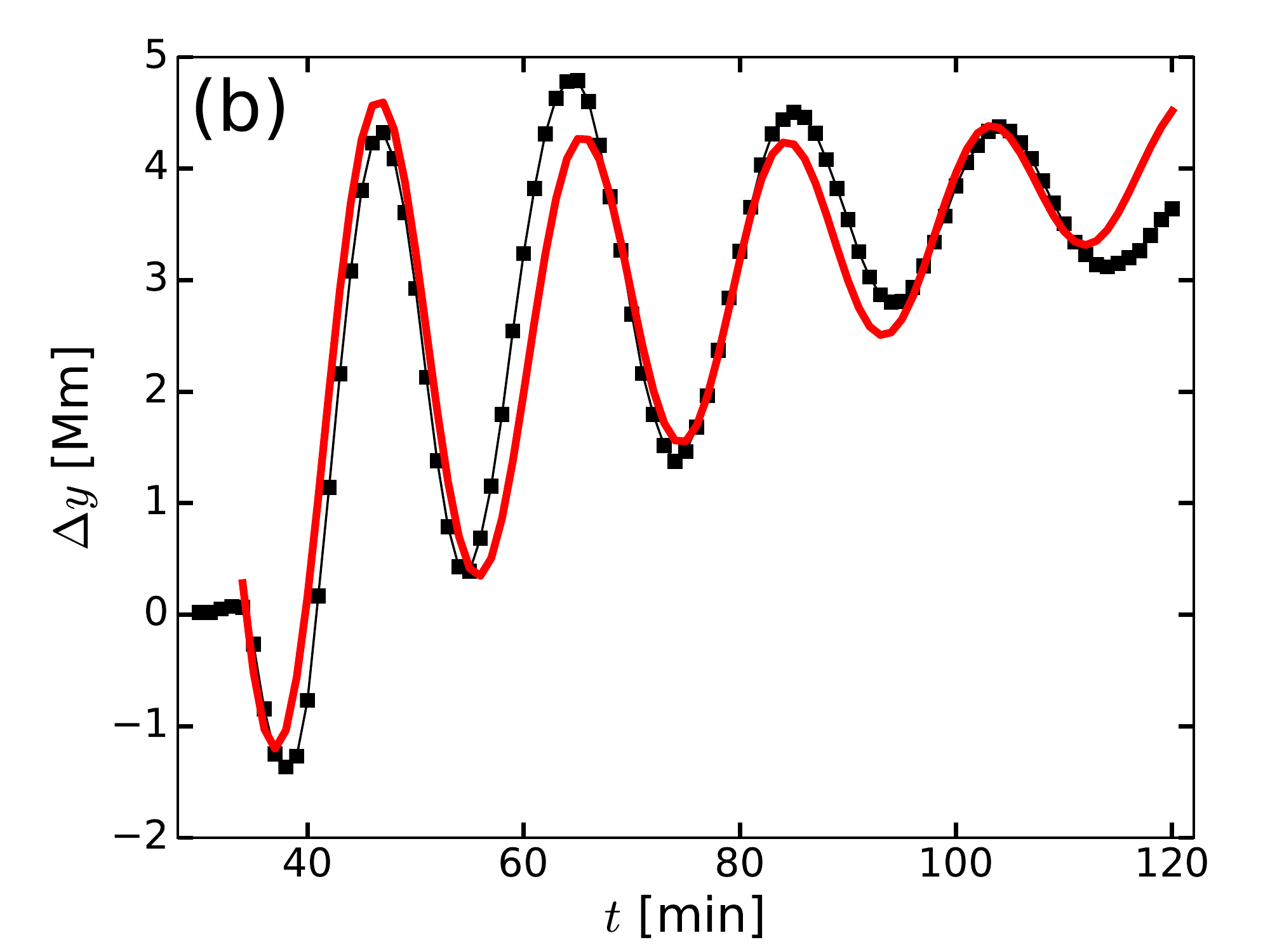}}
\centerline{
\hspace*{0.015\textwidth}
\includegraphics[width=0.515\textwidth]{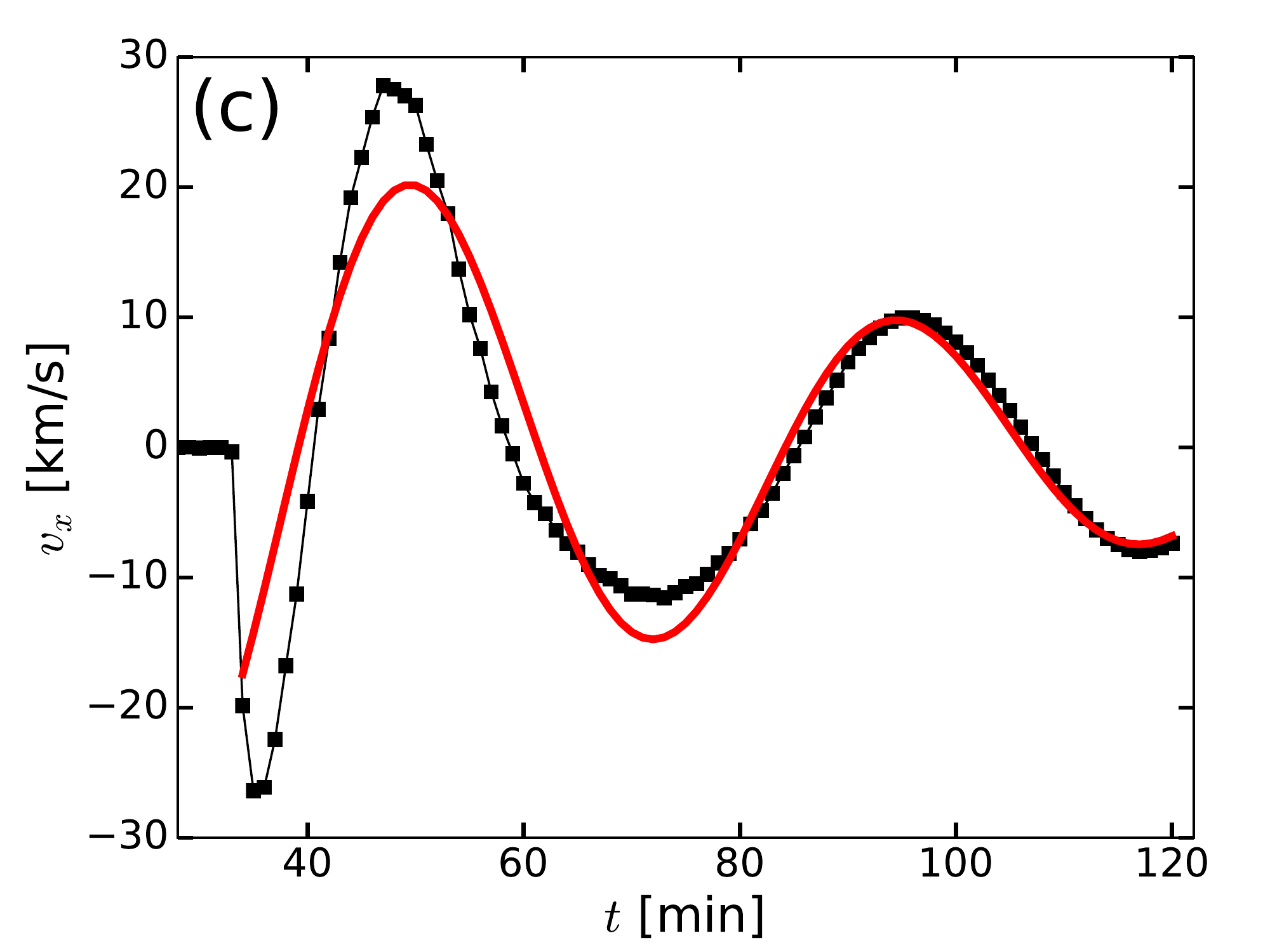}
\hspace*{-0.03\textwidth}
\includegraphics[width=0.515\textwidth]{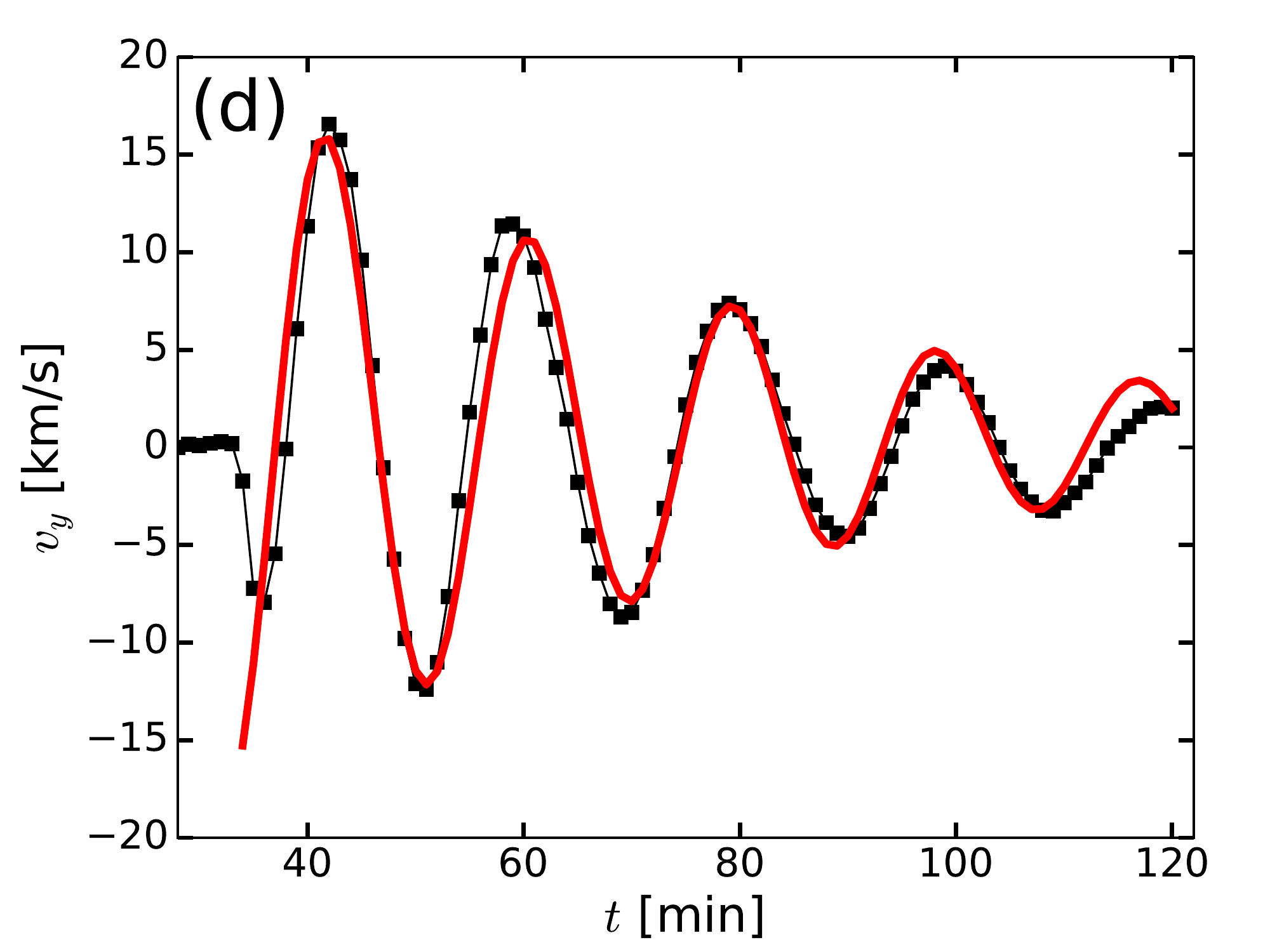}}
\caption{The filament oscillations for Case 3. \textsf{(a)}--\textsf{(b)} Horizontal and vertical displacements are shown as a function of time, just before and after the interaction was triggered: black-squared lines represents the simulation data while red lines represent the fitted curves. Note the displacements are relative to the initial position. \textsf{(c)}--\textsf{(d)} Displayed corresponding horizontal and vertical oscillation speeds.}
\label{f:disp_osc}
\end{figure}

To analyse the filament oscillatory motion, the displacement and velocity of the centre of mass were measured as a function of time for each case. The horizontal and vertical components are represented by $\Delta x$, $\Delta y$, $v_x$ and $v_y$, respectively. Afterwards, the displacements were fitted using an exponential decayed harmonic function $f_i(t)=A_i\sin(\frac{2\pi (t-t_f)}{P_i} +\phi)\exp(-\frac{t-t_f}{\tau_i})+C_i(t-t_f)+f_{0i}$, where $f_i$ holds for the displacements $\Delta x$ and $\Delta y$, $A_i$ is the amplitude of the oscillation, $P_i$ the period, $\tau_i$ the decay constant, $C_i$ the slope, $f_{0i}$ the initial position and $t_f=35~$min the initial fitted time. The index $i$ indicates the fitted variable. The results obtained for all cases are exhibited in Table~\ref{t:fits} (Fitted Parameters). 

As an example, Figure~\ref{f:disp_osc} shows in black-squared lines the displacements and speeds measured for Case 3 along with the fitted curves in red lines. The speed fits are obtained by the time derivative of the displacement fitted functions d$f_i(t)$/d$t$, however the corresponding speed amplitudes displayed in Table~\ref{t:fits} are calculated as the half difference between the minimum and maximum values. The fits show that there is a main frequency describing the oscillation with a certain degree of accuracy. Note that the displacements and speeds are almost null before the interaction begins and then show a damped oscillation with the vertical motion exhibiting a small upward slope. This upward slope is mainly due to a slight reduction of the filament density with time because of numerical diffusion. Comparing the left panels \textsf{(a)}--\textsf{(c)} with the right ones \textsf{(b)}--\textsf{(d)}, it is clearly seen that the horizontal period is larger than the vertical one. In the same way, the horizontal damping time is larger than the vertical one (see Table~\ref{t:fits}). 
This behaviour is qualitatively similar for all cases.

\begin{figure}
\centerline{
\hspace*{-0.02\textwidth}
\includegraphics[width=0.53\textwidth]{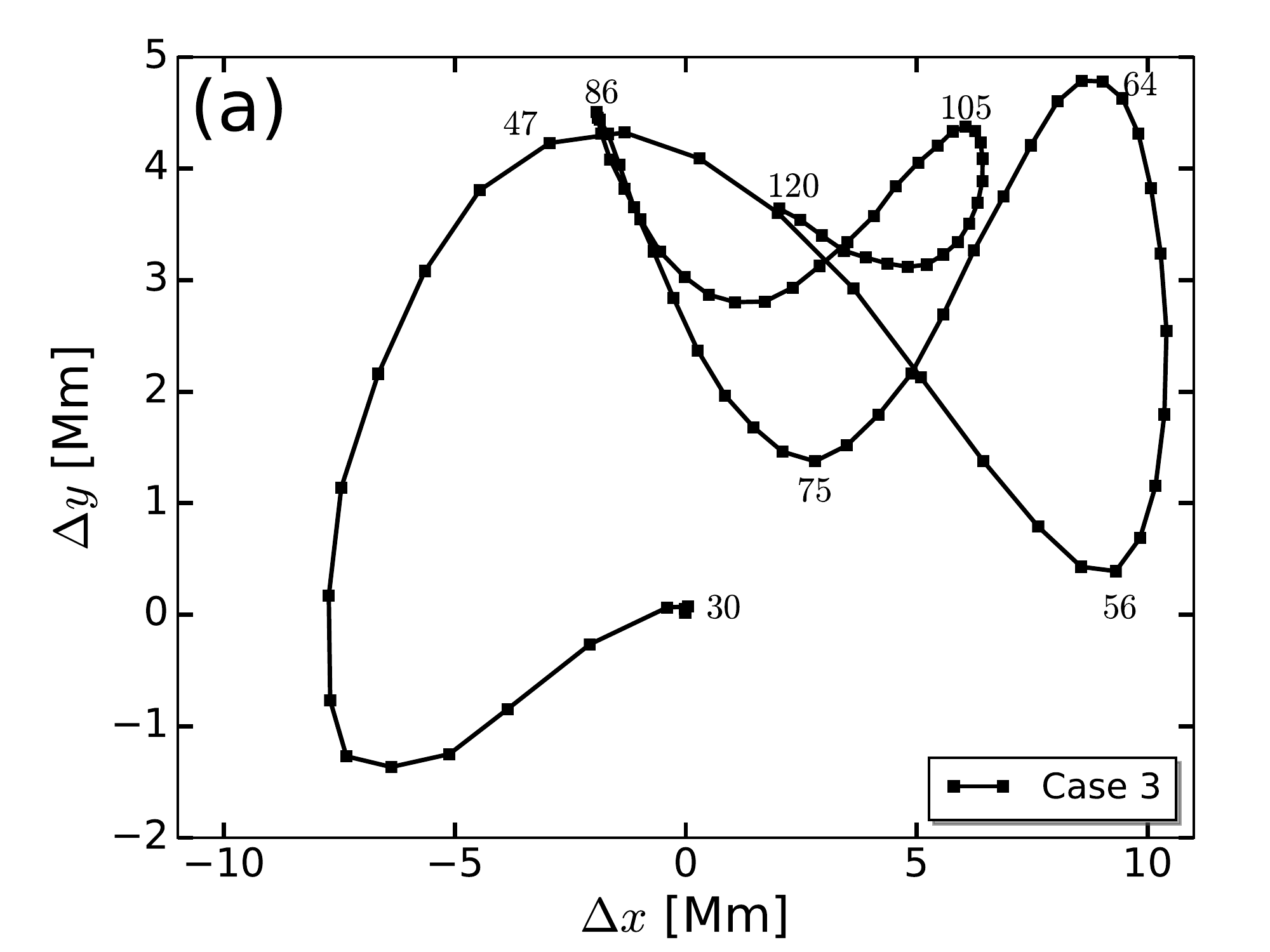}
\hspace*{-0.04\textwidth}
\includegraphics[width=0.53\textwidth]{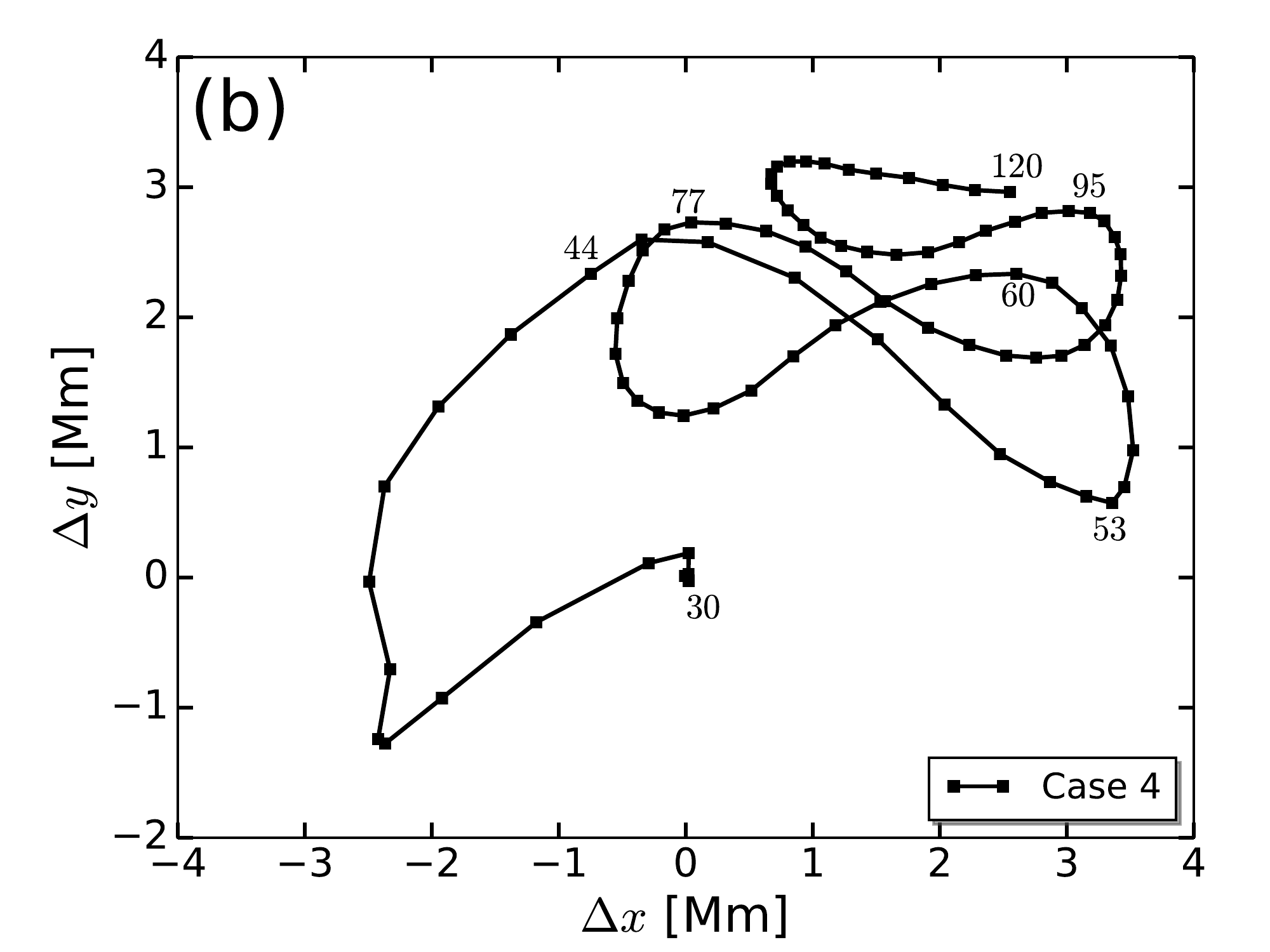}}
\caption{Filament trajectories during the oscillating stage corresponding to Case 3 (panel \textsf{(a)}) and Case 4 (panel \textsf{(b)}). The displacements are referred to the initial positions $(x,y)=(0,h)~$Mm. The sequential numeric annotations indicate the times (in min) corresponding to those positions helping to understand the temporal evolution.}
\label{f:traj}
\end{figure}

Figure~\ref{f:traj} displays the trajectories for Cases 3 and 4 during the oscillating stage. As the displacements are referred to the initial position, both trajectories almost begin at the origin indicated by their corresponding times $t=30~$ min. The numeric points are time sequenced to indicate the temporal evolution. Comparing the trajectories with those determined analytically by \citet[][see their figures 3 and 4]{2018kolotkovJASTP172} utilising a similar model as ours, we can see that the trajectories in Figure~\ref{f:traj} resemble Lissajous-like curves of an hourglass shape suggesting that the  coupled transverse oscillations are close to a resonant regimes. Note the patterns shown in Figure~\ref{f:traj} are distorted from the symmetric Lissajous-like curves because the filament is finite; it oscillates embedded in an inhomogeneous plasma; and due to the action of dissipative effects (numerical diffusion and damping, see Section \ref{Sub:diff}). Not all simulated cases exhibit a Lissajous-like pattern, specially Case 5 which is strongly damped.

\subsubsection{Numerical Diffusion and Damping}
\label{Sub:diff}
It is important to verify that the filament attenuated oscillations (e.g., as can be appreciated from Figure~\ref{f:disp_osc} for Case 3) are not due to the artificial viscosity produced by  truncation errors of the numerical scheme used, whose magnitude depends on the order of accuracy of the method and is generally reduced by an increase in the grid resolution. However, a direct evaluation of the numerical diffusion is not a simple task because the modified set of Equations \ref{e:mhd} (i.e., the governing equations affected by the truncation error terms of the numerical method used) is too hard to obtain for high-order MUSCL type schemes. A numerical estimation is also difficult  since the diffusion effects are different along the domain due to gradient changes and variations of the spatial resolutions when using AMR meshes.

\begin{figure} 
\centerline{
\hspace*{-0.02\textwidth}
\includegraphics[width=0.52\textwidth]{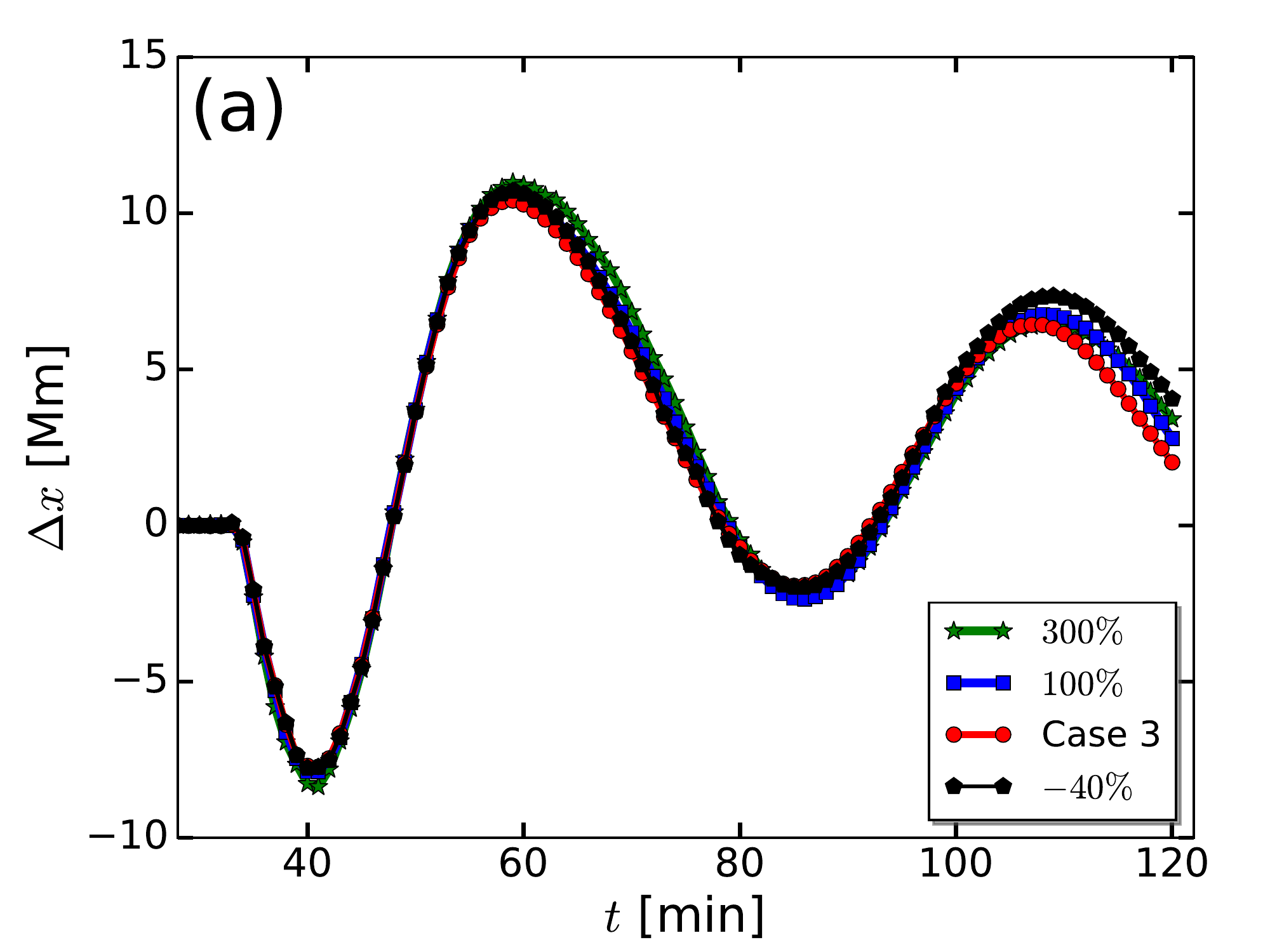}
\hspace*{-0.04\textwidth}
\includegraphics[width=0.52\textwidth]{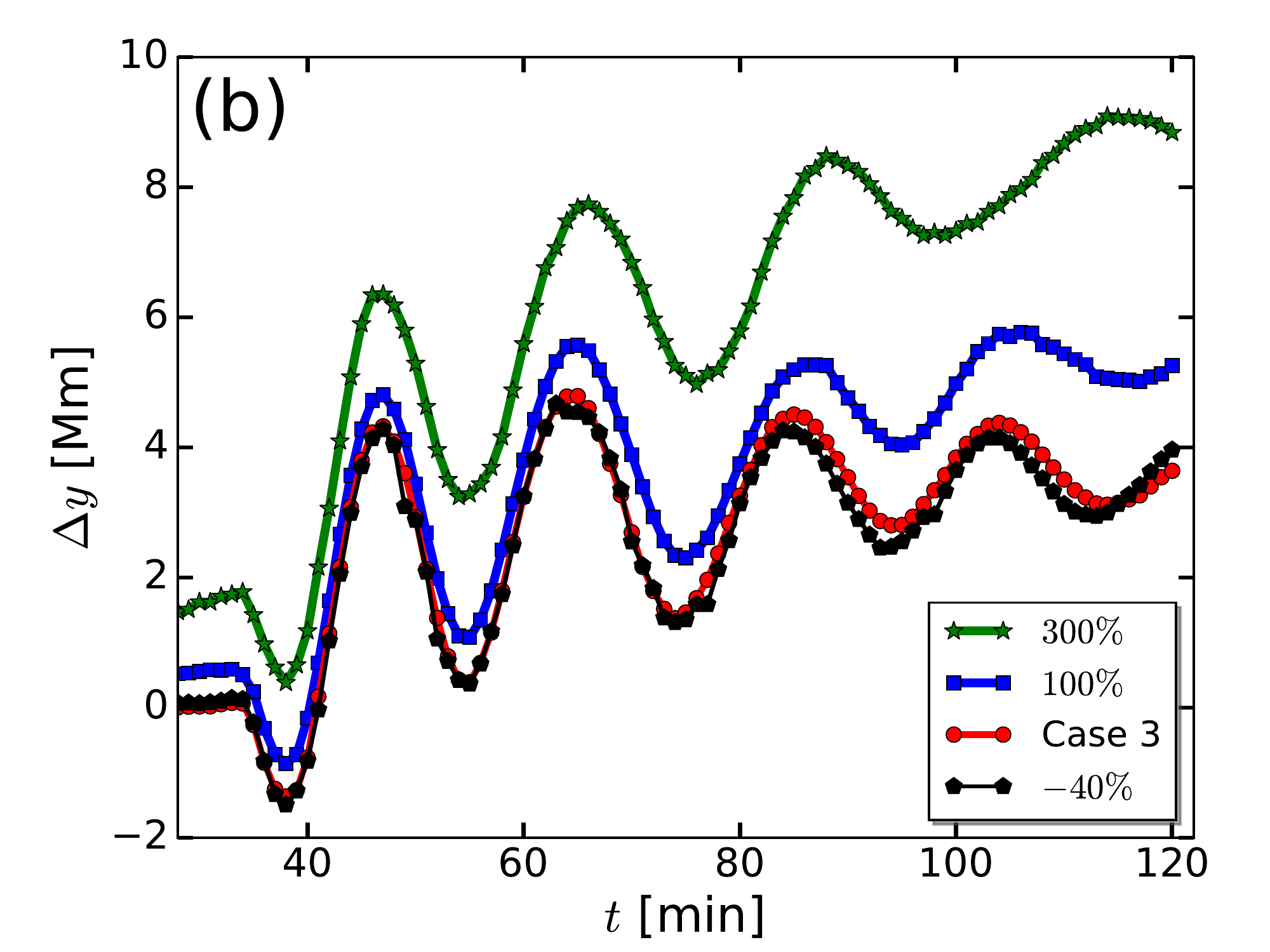}}
\caption{Test to determine weather the damped motion exhibited by the filament is due to numerical diffusion or it has a physical origin. Four runs were performed varying the spatial resolution, corresponding to relative coarser grids of $300\%$, $100\%$ with respect to the standard resolution of Case 3, and a relative finer grid of $-40\%$. We concluded that the acting  damping process has a significant physical meaning. \textsf{(a)--\textsf{(b)}} Show horizontal and vertical displacements for the different resolutions.} 
\label{f:resol} 
\end{figure}

For the analysis we chose Case 3 as a reference run and performed simulations with different spatial resolutions to evaluate changes in the numerical results. If the damping time is a consequence of the numerical diffusion, it will be shorter for coarser grids because the numerical diffusion coefficient increases when the grid resolution is reduced. This means that the filament motion should be damped in fewer cycles for coarser grids than for finer ones. Figure~\ref{f:resol} displays the numerical results for the displacements using four different progressively refined grids ($\delta x=\delta y=[0.37, 0.19, 0.094, 0.059]~$Mm), corresponding to relative coarser grids of $300\%$, $100\%$ and $0\%$, with respect to the standard resolution of Case 3 (see Section \ref{Sub:code}), and a relative finer grid of $-40\%$. The AMR meshes used for the runs $300\%$, $100\%$ and Case 3 were of six, seven and eight levels of refinement, respectively, with $20\times 10$ cells per block; whereas for the run $-40\%$ was of eight levels with $32\times 16$ cells per block. Taking into account that the upward slopes exhibited by the vertical displacements $\Delta y$, which are smaller for higher grid resolutions, are due to the numerical imbalances between the filament and its background, not affecting the oscillatory part of the motion, we concluded that the numerical results of Case 3 are not significantly influenced by the artificial viscosity. In fact, we can see that there is no direct correlation between the damping level and the grid resolution since, for example, the motion in the $x$-axis for the reference run damps a bit faster than for the coarser ones. Therefore, for all studied cases using the reference grid resolution, we can assume that the numerical diffusion has a minimum impact on the damping process and can be reasonably ignored.

With regards to the filament attenuated oscillations, since we are not considering non-adiabatic processes (as radiative losses or heat conduction) and according to the geometry of the problem, we discuss two main possible damping mechanisms: resonant absorption and wave leakage. 

\begin{figure}
\centerline{
\hspace*{-0.02\textwidth}
\includegraphics[width=0.49\textwidth]{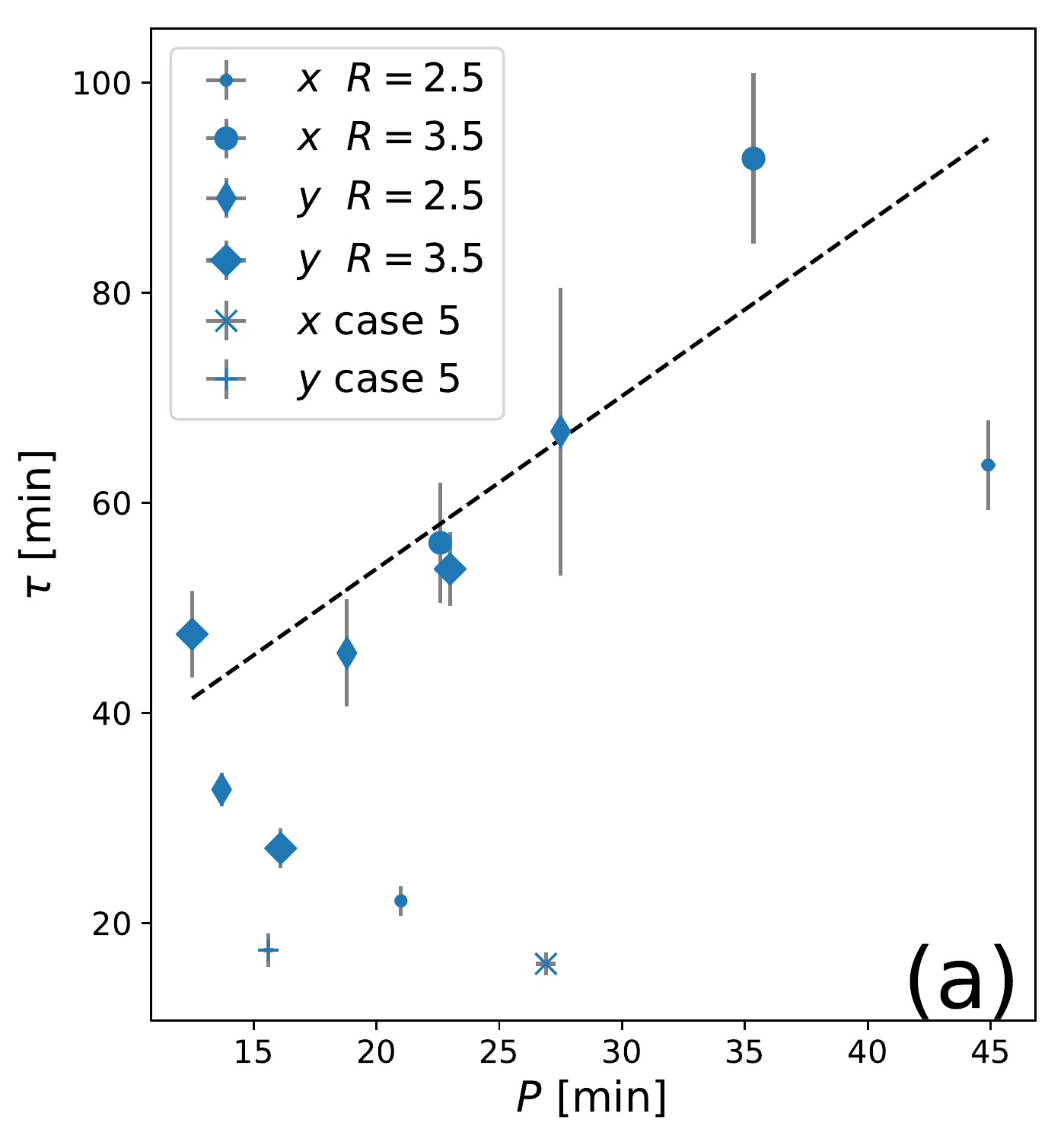}
\hspace*{-0.02\textwidth}
\includegraphics[width=0.5265\textwidth]{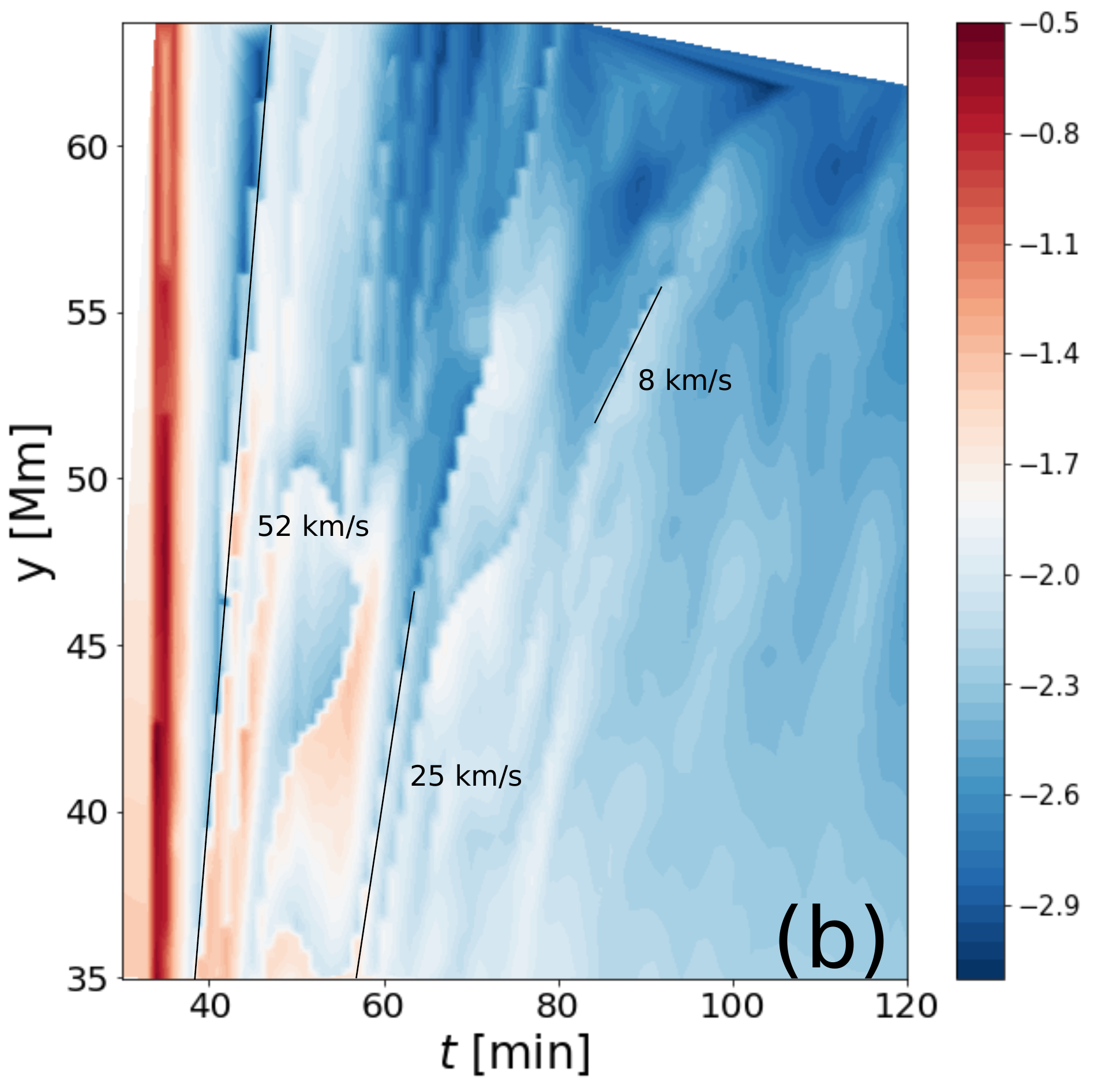}}
\caption{\textsf{(a)} Damping times vs periods for the cases exhibited in Table~\ref{t:fits} except Cases 6 and 7. The values for the horizontal (vertical) direction are displayed with circle (diamond) symbols. Case 5 is displayed with  cross  symbols. The widths of symbols are according to the radius of the filament. The grey lines correspond to the errors of the variables. The black-dashed line shows a linear fit. \textsf{(b)} Pressure stack plot of Case 3 along the vertical dash line in Figure~\ref{f:pres_patt}\textsf{(d)}. The black solid lines outline the speed of slow waves propagating away from the filament. The colour map represent $ln(p)$}
\label{f:damping}
\end{figure}

First, we consider the resonant absorption of infinitely long thread oscillations \citep[see e.g.,][and references therein]{2018arregieLRSP15}. This mechanism relies on the energy transfer from transverse kink modes to small scale Alfv\'en waves because of the plasma inhomogeneity located at the transition layer between the filament and the surrounding corona. Assuming a small transition layer ($\Delta/R <<1$), analytical expressions for the damping time scale were obtained for the kink modes \citep[see e.g.,][]{1992SoPh..138..233G,2002A&A...394L..39G,2002ApJ...577..475R}, claiming a linear dependence between periods and damping times. In order to explore a possible relation, we fit a linear function $\tau=aP+b$ using the values in Table \ref{t:fits}. Figure~\ref{f:damping}\textsf{(a)} shows the damping times against the periods for horizontal (circles) and vertical (diamond) oscillations. Case 5 is pointed out with a cross and an asterisk. The error bars are in grey colour, for the periods these errors are comparable to the symbol size. Small circles and diamonds correspond to cases with filament radius $R=2.5~$Mm and big symbols to $R=3.5~$Mm. The black-dashed line displays the linear fit with $a=(1.6\pm0.5)$ and $b=(20.8\pm 13.7)~$min. For the analysis we excluded the damping times of Cases 6 and 7 because they are out of range. This crescent linear tendency looks similar to the behaviour  by \citet{2011A&A...531A..53H} suggesting that the resonant absorption might be acting as a damping mechanism. In addition as a complementary study, taking into account the work of \citet{2016ApJ...820..125T} in which they found an energy enhancement in a thin region near the filament's edge attributed to resonant absorption, we explored the evolution of the kinetic energy in a shell surrounding the filament's core ($R_2<r<5~ {\rm Mm}$). The results did not show a definite  kinetic energy enhancement associated with  the surrounding shell to support the linear tendency found in Figure~\ref{f:damping}\textsf{(a)}. 
Thus, more studies are required to investigate  to what extent the resonant absorption may be responsible for the oscillation damping.  

Second, to study weather the wave-fronts observed emanating from the filament (such as in Figures \ref{f:pres_patt}\textsf{(c)}--\textsf{(d)}) are evidence of wave leakage, we constructed for Case 3 a pressure stack plot for each time measured along a vertical slice located at $(x,y)=(0,35$--$65)~$Mm (highlighted in Figure~\ref{f:pres_patt}\textsf{(c)} by a dash line). The stack plot results are shown in Figure~\ref{f:damping}\textsf{(b)} where wave signatures propagating with speeds of $\sim$52~km s$^{-1}$ at an early time, and later with $\sim$25~km s$^{-1}$ and $\sim$8~km s$^{-1}$ can be appreciated. Also, the periods of these waves, which can be deduced from the separation in time between the fronts, is of $\sim$20~min, hence associated with the periods of the filament oscillation. On the slice domain the spatial average sound speed $\bar{c}_{\rm s}=150~$km s$^{-1}$, the average Alfv\'en speed $\bar{v}_{\rm A}=185~$km s$^{-1}$ and the average plasma $\bar{\beta}=0.94$. From Figures~\ref{f:pres_patt}\textsf{(c)}--\textsf{(d)} and \ref{f:damping}\textsf{(b)} we identify compressive, slow magneto-acoustic waves emanating from the filament with an oblique propagation angle with respect to the magnetic field, which seems to be the reason why the wave-front speeds are markedly smaller than the sound speed. Thus it seems that, while the filament oscillates, it acts as a forcing driver transferring energy into the surrounfing plasma. On the other hand, during the oscillating stage the filament lost into the environment $\sim$3$\%$ of its mass. We thus conclude that the main process responsible of the oscillation damping seems to be the wave leakage.

\begin{table}
\caption{Set-up and fitted parameters for all numerical cases. Varying parameters: the initial filament height $h$, filament radius $R$, azimuthal current density $j_1$ and maximum simulation time $t_{\rm {max}}$. $\bar{\rho}_{\rm fil}$ is the averaged filament density. Fix parameters for all cases: the axial current density $j_0$, density at the coronal base $\rho_0$, filament transition layer $\Delta$, dipole intensity $M_{\rm{dip}}$, coronal temperature $T_0$, chromospheric temperature $T_{\rm cr}$, filament temperature $T_{\rm{fil}}$, chromospheric width $h_{\rm{cr}}$ and transition region width $h_{\rm{tr}}$. Fitted parameters in the horizontal $x$-axis: the period $P_x$, displacement amplitude $A_x$, the damping time $\tau_x$ and the speed amplitude $A_{v_x}$. Same is displayed for the vertical $y$-axis.}
\label{t:fits}
\begin{tabular}{lccccccc}  
\hline                     
                           & \tiny{Case 1} & \tiny{Case 2} & \tiny{Case 3} & \tiny{Case 4} & \tiny{Case 5} & \tiny{Case 6} & \tiny{Case 7}\\ 
\hline 
\underline{Set-up parameters}&      &       &        &         &        &        &        \\
$h~$[Mm]                   &  14.   &  14.  &   19.  &   19.   &  19.   &  25.   &  25.   \\
$R~$[Mm]                   &   2.5  &  3.5  &   2.5  &   3.5   &  2.5   &   2.5  &  3.5   \\
$j_1~$[statA cm$^{-2}$]    &  600.  &  600. &   600. &   600.  & 797.5  & 600.   &  600.  \\
$t_{\rm {max}}~$[min]      & 120.   & 120.  & 120.   & 120.    & 120.   & 150.   & 150.   \\  
$\bar{\rho}_{\rm fil}~$[g cm$^{-3}\times 10^{-13}$] &4.0&7.2&4.0&7.2&2.5&4.0    &7.1     \\
$j_0~$[statA cm$^{-2}$]    & \multicolumn7c{600.}             \\ 
$\rho_0~$[g cm$^{-3}$]    & \multicolumn7c{$3.2\times 10^{-15}$} \\
$\Delta~$[Mm]              & \multicolumn7c{1.25}             \\
$M_{\rm{dip}}$             & \multicolumn7c{$2.25$}           \\ 
$T_0~$[MK]                 & \multicolumn7c{1.}               \\
$T_{\rm cr}~$[K]           & \multicolumn7c{$1.2\times 10^4$} \\
$T_{\rm{fil}}~$[MK]        & \multicolumn7c{0.3}              \\
$h_{\rm{cr}}~$[Mm]         & \multicolumn7c{2.5}              \\
$h_{\rm{tr}}~$[Mm]         & \multicolumn7c{0.5}              \\
\hline                                  
\underline{Fitted parameters}     &        &      &          &        &        &        &        \\ 
                                  &        &      &          &\underline{$x$-axis} &  &     &    \\
$P_x~$[min]                       & 21.0    & 22.6 & 44.9     & 35.3   & 26.9   & 66.9   & 51.7  \\
$A_x~$[Mm]                        & 6.0     & 1.5  & 11.2     & 3.1    & 15.2   & 22.2   & 6.7   \\
$\tau_x~$[min]                    & 22.1   & 56.2 & 63.6      & 92.8   & 16.1   & 197.2  & 354.4 \\
$A_{v_x}~$[km s$^{-1}$]           & 19.8   & 7.2  & 27.1     & 9.9   & 36.0   & 29.9   & 13.5    \\  
                                  &        &      &          &\underline{$y$-axis} &  &  &       \\
$P_y~$[min]                       & 13.7   & 12.5 & 18.8     & 16.1   & 15.6   & 27.5   & 23.0   \\  
$A_y~$[Mm]                        & 2.7    & 1.4  & 3.3      & 2.2    & 5.1    & 2.6    & 2.9    \\  
$\tau_y~$[min]                    & 32.7   & 47.5 & 45.7     & 27.1   & 17.4   & 66.8   & 53.7   \\
$A_{v_y}~$[km s$^{-1}$]           & 15.6   & 8.9 & 14.5     & 9.1   & 18.7   & 6.7    & 9.7      \\  
\hline
\end{tabular}
\end{table}

\subsubsection{Restoring Forces}
\label{Sub:force}

\begin{figure} 
\centerline{
\hspace*{-0.01\textwidth}
\includegraphics[width=0.53\textwidth]{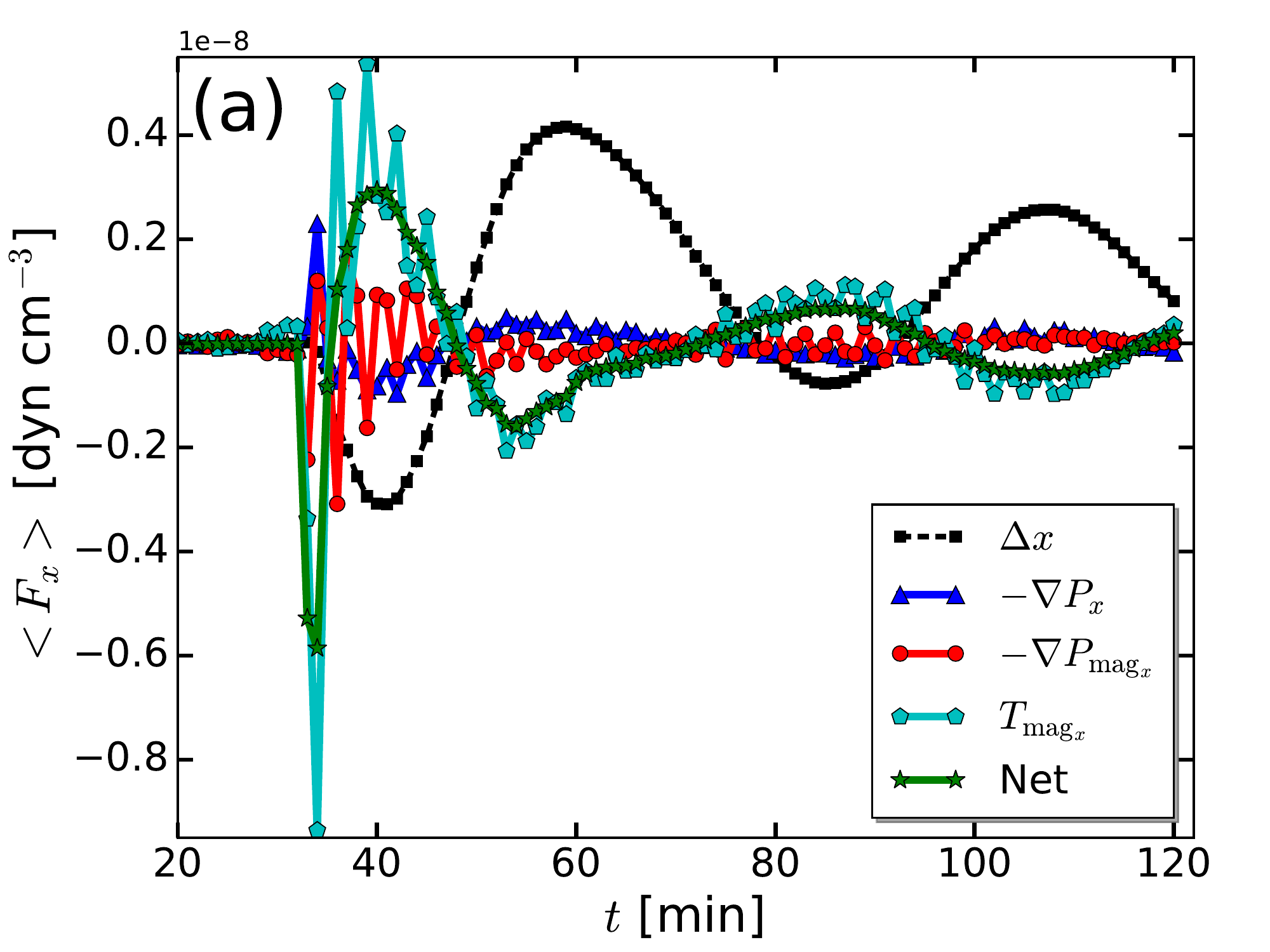}
\hspace*{-0.04\textwidth}
\includegraphics[width=0.53\textwidth]{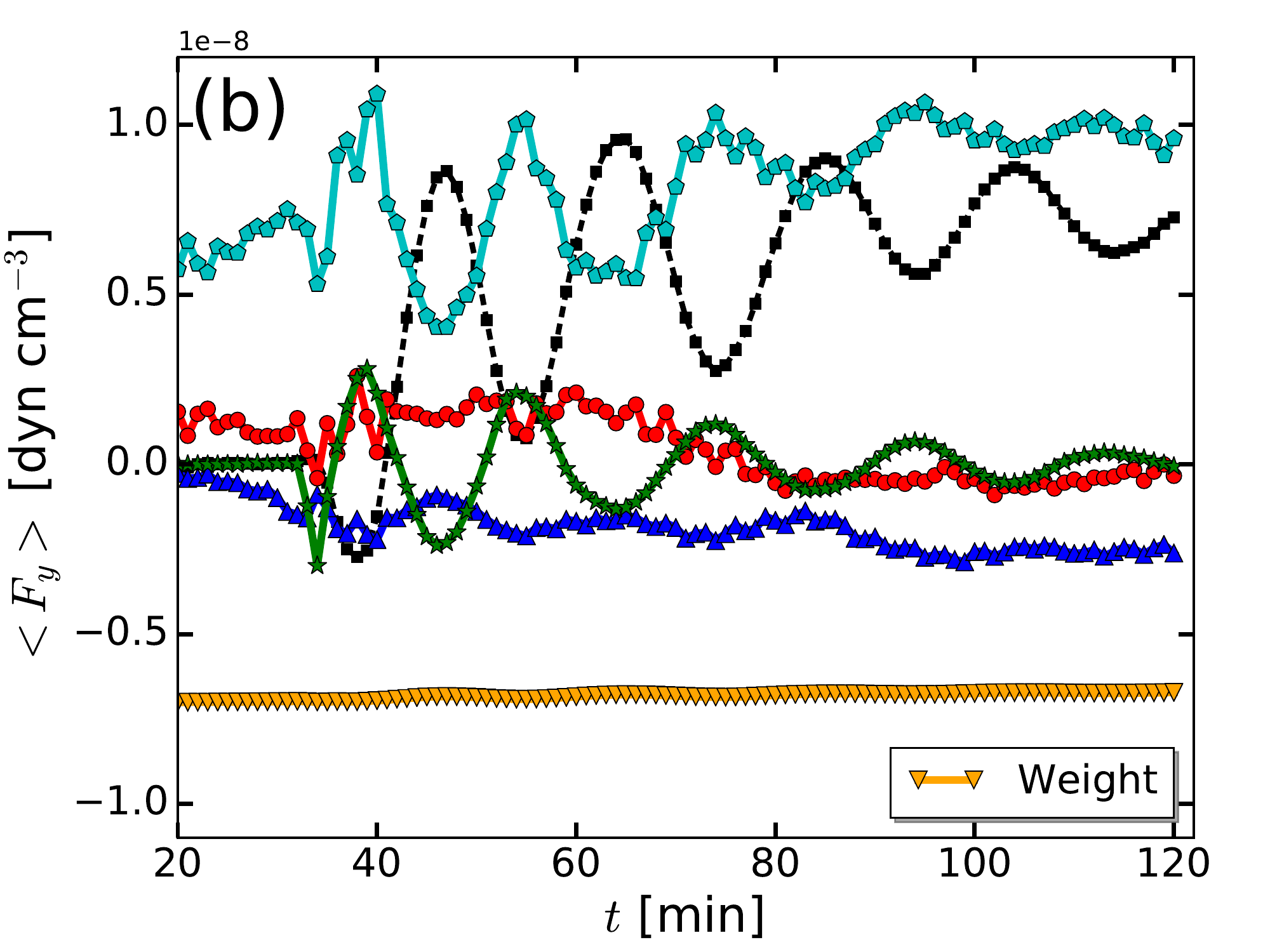}}
\caption{Forces acting on the oscillating filament. \textsf{(a)} Averaged forces per unit  volume in the horizontal direction:  a reference square-black-dashed line depicts the displacement with its scale modified;   triangle-up-blue-solid line represents the thermal pressure; circle-red-solid line the magnetic pressure; pentagon-cyan-solid line the magnetic tension and asterisk-green-solid curve the net force. \textsf{(b)} Similar to panel (a) but corresponding to the vertical component: square-black-dashed line represents vertical displacement with a modified scale; triangle down-orange-solid line is the almost constant weight force.} 
\label{f:fuerzas}
\end{figure}

Figures \ref{f:fuerzas}\textsf{(a)}--\textsf{(b)} display the calculation of the thermal pressure force $-\mathbfit{\nabla}p$ (blue lines) and the Lorentz force $\frac{1}{c}\mathbfit{j}\times\mathbfit{B}$, decomposed into the magnetic tension $\frac{\mathbfit{B}\cdot\mathbfit{\nabla}\mathbfit{B}}{4\pi}$ (cyan lines) and magnetic pressure force $-\mathbfit{\nabla}(\frac{B^2}{8\pi})$ (red lines), along the horizontal and vertical axes, respectively. In green lines we represented the net forces. Also, as a reference, in black lines the filament displacements were added with an appropriate scale, to facilitate a  visual comparison. The Figure~\ref{f:fuerzas} corresponds to Case 3 and displays the averaged forces over the grid-cells belonging to the filament structure ($r<R_3$) evolving in time. With respect to the horizontal direction, panel \textsf{(a)} shows that,  before the interaction begins (by $t<33~$min), all forces are almost null, as expected. Then the forces behave noisily during a very short time lapse, attributed to the strong interaction with the shock waves. Later on,   the magnetic tension results the main restoring force (more smoothly drawn by the net force in green colour). Meanwhile, for early times, the thermal pressure force seems to follow the displacement and the magnetic pressure one does not play a significant role. Regarding the forces acting on the vertical direction, panel \textsf{(b)} shows that the magnetic tension is always positive, pointing upwards, opposing  the almost constant and negative weight force (orange line). It is noticeable the correspondence between the magnetic tension and the vertical displacement. Note that the main restoring force in the upward direction is the magnetic tension, while in the downward direction is the weight. In turn, the thermal and magnetic pressure forces play secondary roles, the thermal pressure force becomes negative for later times, while the magnetic pressure forces pushes upwards until finally vanishes, by $t\approx 80~$min.

\subsection{Parametric study} 
\begin{figure}
\centerline{
\hspace*{0.0\textwidth}
\includegraphics[width=0.51\textwidth]{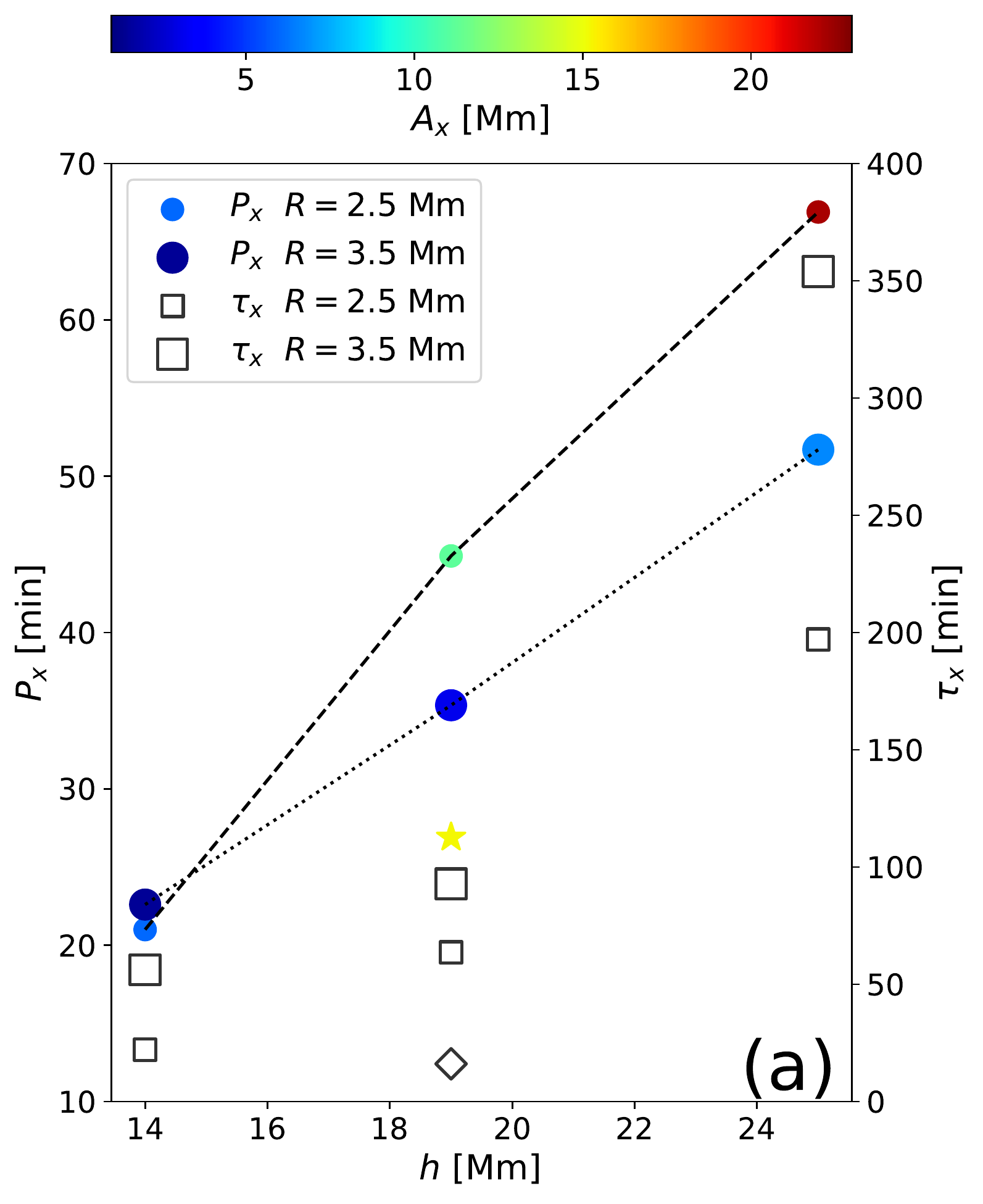}
\hspace*{-0.02\textwidth}
\includegraphics[width=0.52\textwidth]{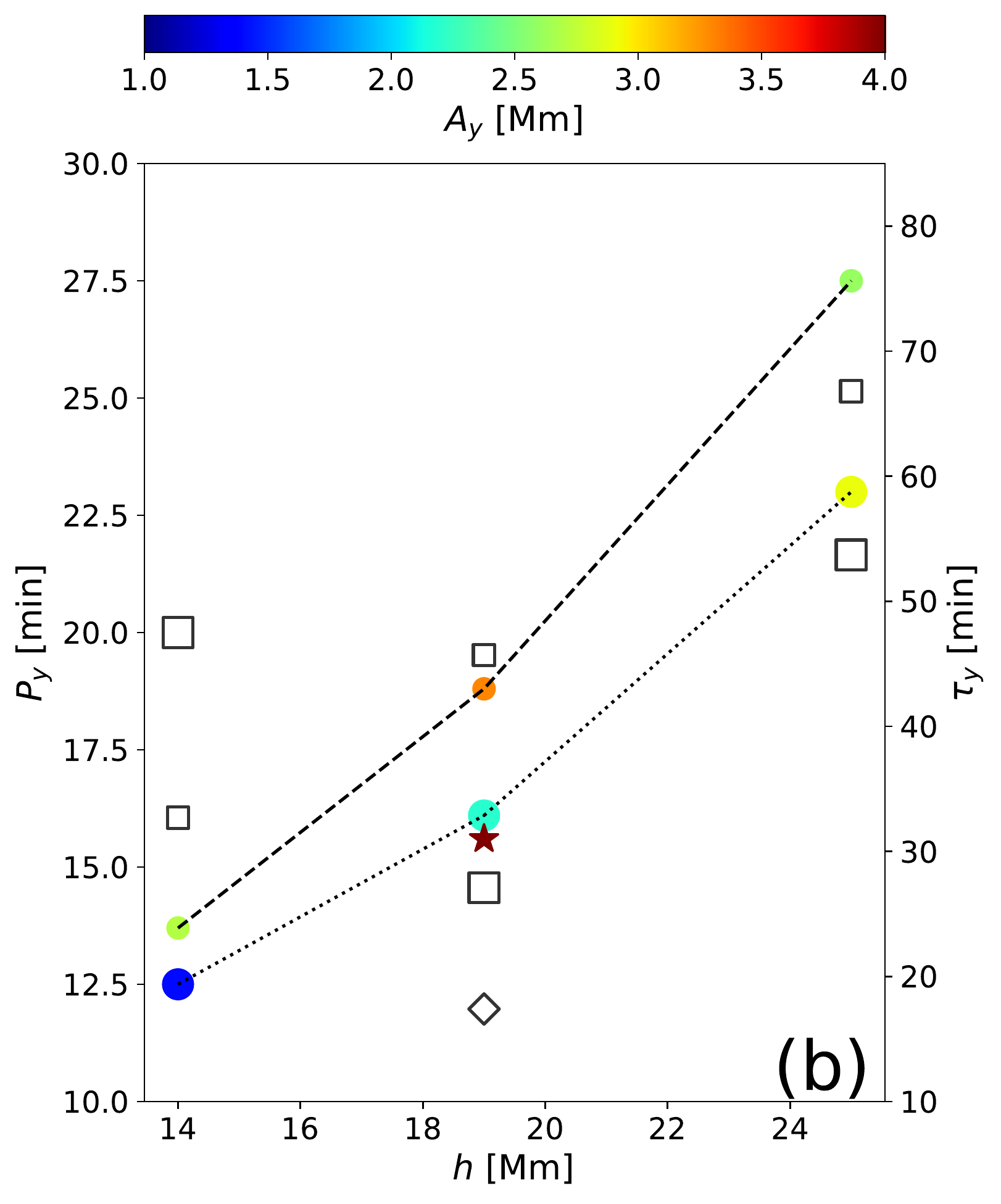}} 
\caption{Fitted parameters as a function of height for all cases. Panels \textsf{(a)}--\textsf{(b)} show periods, amplitudes of oscillation and damping times obtained for the horizontal and vertical displacements, respectively. The circles indicate the periods  and the colour map exhibits the amplitudes. Small circles connected with a dashed line represent the cases with lower filament radius, and big circles, joined by a dotted line, identify  cases with larger radius. The squares display the damping times. The squared symbol sizes follow the same rules as the circles. The period for Case 5 is pointed out with a star and the damping time with a diamond symbol.}
\label{f:results}
\end{figure}

Figure~\ref{f:results} summarised the fitted parameters of all cases for the horizontal and vertical displacements (also listed in Table~\ref{t:fits}). Left (right) panel corresponds to the  $x$-($y$-)axis. In each panel, circles indicate periods as a function of height, and the colour map describes the amplitudes of oscillation. Small circles represent the triad of Cases 1--3--6 with filament small radius $R=2.5~$Mm, connected by dashed lines; while the big circles indicate the triad of Cases 2--4--7 with large radius $R=3.5~$Mm, linked by dotted lines. We also show the damping times, depicted by squared symbols following equal logic. Case 5, with a different filament mass, is highlighted with different symbols, a star for the period and a diamond for the damping time.

In the following we itemise the main trends obtained for the seven cases. Beginning with the $x$-direction represented by Figure~\ref{f:results}\textsf{(a)}:
\noindent
\begin{enumerate}
\item Comparing cases of same height but different radii: large radii are generally associated with smaller periods, smaller amplitudes (darker blue) and larger damping times than small radius cases. This tendency is seen comparing the pairs of Cases 1--2, 3--4, and 6--7, except for the pair 1--2 of similar periods.
\item Comparing cases of same radius but different heights: larger heights are associated with larger periods, larger amplitudes and larger damping times. This occurs for the triads of Cases 1--3--6 and 2--4--7. 
\item Comparing cases of same radius and height but different filament masses: Case 5 shows a smaller period, a larger amplitude and a considerable smaller damping time than the heavier Case 3.
\end{enumerate}
With respect to the $y$-direction (see Figure~\ref{f:results}\textsf{(b)}), the fitted parameters show similar trends as just described for the $x$-direction but having a less regular behaviour:
\begin{enumerate}
\item Comparing cases of same height but different radii: the pairs of Cases 1--2, 3--4 and 6--7 show that the increase of the radius implies smaller periods, smaller amplitudes (except the pair 6--7) and smaller damping times (except the pair 1--2).
\item Comparing  cases of same radius but different heights: the triads of Cases 1--3--6 and 2--4--7 show that the increase in height commonly denotes larger periods, larger amplitudes (except the pair 3--6) and larger damping times (except the pair 2--4).
\item Comparing cases of same radius and height but different filament masses: Case 5 exhibits a smaller period, a larger amplitude and a quite smaller damping time than the heavier Case 3. 
\end{enumerate}
Also, comparing the results between the horizontal and vertical oscillations, we note that the horizontal amplitudes, periods and damping times are larger than the vertical ones, in agreement with  \citet{1999A&A...345.1038S}.

Summarising, there is a general tendency of periods, displacement amplitudes and damping times to increase with height. This tendency is clearly observed for horizontal oscillations, although it is not always valid for vertical ones. The same behaviour is followed by the velocity amplitudes. The general tendency to increase with height is in agreement with the fact that the magnitude of the net force acting on the filament, along with its restoring frequency, diminish with height (the net force frequency can be appreciate, e.g., in Figure~\ref{f:fuerzas}).

To increase the radius in general implies shorter periods and smaller amplitudes. But, while it clearly implies longer damping times considering the horizontal oscillations, we found shorter damping times when considering  vertical oscillations. Note that the increase of  the radius produces an increase on the net force frequency acting over the filament, which explains the shorter periods found. 
Finally, exploring variations of the filament mass, while keeping fixed the remaining independent  parameters of the set-up, the lower mass Case 5 exhibits shorter periods and larger amplitudes in both axes, although more noticeable is that Case 5 shows a stronger damped motion. 

As  exposed above, the horizontal and vertical oscillation properties of the filament are, in general, different. Thus, in observations using the Doppler technique to measure oscillation properties, projection effects along the line-of-sight may play a significant role in the results. For instance, in the framework drawn in Figure~\ref{f:esquema} a particular observer whose line-of-sight is that of a prominence seen at the limb ($\phi=0$) will measure the horizontal component of the oscillation; whereas, an observer with a top-down view of a filament at the centre of the solar disc ($\phi=90$) will measure the vertical component. The Doppler velocity measured by intermediate observers will correspond to the filament velocity vector projected along the line-of-sight of the observer. Therefore, as the winking phenomenon related to the Doppler shift is more likely to be detected in filaments with large velocity amplitudes, in the framework of our results this will correspond to Cases $[1,3,5,6]$ and with line-of-sights near the limb, that is when the horizontal velocity component, which is frequently larger than the vertical one, contributes more to the Doppler effect.

\section{Conclusions}
\label{S-Conclusion}
In order to contribute to the understanding of the oscillatory phenomenon of quiescent filaments, we performed a parametric study of large-amplitude transverse oscillations carrying out 2.5D ideal MHD simulations. The model of \citet{1990forbesJGR95} was adapted to numerically represent  a filament floating in a gravitationally stratified atmosphere. After the set-up initialisation and a relaxation process, the quiescent filament was perturbed using a blast device that excites coupled horizontal and vertical oscillations for a few cycles. 

By analysing the forces, even though our model differs from other authors, either in the magnetic configuration (Kippenhan and Schl\"uter or Kuperus and Raadu) or in the perturbation mechanism, we also found that the main restoring force is the magnetic tension. As suggested by \citet{2014shenApJ795}, along the vertical direction the main restoring force is the coupling between the gravity and magnetic tension. Whereas, along the horizontal direction the main  force is  the magnetic tension. In agreement with \citet{2018zhouApJ856} we found that the magnetic pressure  plays a secondary role and  the gas pressure contribution  is negligible. 
 
The oscillation properties present different behaviours along the horizontal and vertical axes. The periods, amplitudes and most of the damping times are larger in the horizontal direction. The filament exhibits a vertical oscillation frequency being approximately twice  larger than the horizontal one (listed in Table~\ref{t:fits}), corresponding to coupled transverse oscillations, which in some cases seem close to a resonant regime (see Figure~\ref{f:traj}) indicated by the resemblance of the trajectory pattern with the Lissajous-like curves obtained by \citet{2018kolotkovJASTP172}. This behaviour may indicate that the vertical motion is a reaction to the horizontal oscillation, as  pointed out by \citet{2020liakhA&A637}.

Also, our results (Cases 1 and 5) are in good agreement with the results in \citet{2009tripathiSSRv149} and \citet{2014shenApJ795} who found a range of periods of 11--29~min, velocity amplitudes of 6--41~km s$^{-1}$, and damping times of 25--180~min. Whereas Cases 6 and 7 keep some agreement with the events studied by \citet{2015RAA....15.1713P}, who reported values of 61--67~min, 12--22~km s$^{-1}$ and 92--117~min, respectively.

In order to explore the role of different parameters on the oscillatory motion, a parametric study was carried out performing seven simulations varying the height, size and mass of the filament. The fitted parameters revealed, as a first general rule, that periods, amplitudes and damping times increase with height (Figure~\ref{f:results}). This tendency is clearly observed in the horizontal direction, but has some exceptions in the vertical one. Given that the net force exerted onto the filament decreases with height and its restoring frequency as well, this results in larger periods and amplitudes measured on higher filaments.

This tendency of periods to increase with height partially agrees with the analytical model of \citet[][see their figure 3]{2016kolotkovA&A590} who considered uncoupled linear oscillations. For both the vertical periods increase with height, whereas for the horizontal periods these authors found a tendency to diminish with height.

Our first general tendency is in agreement with observational cases, for example: \cite{2013franchileAA552} analysed a filament with a low height of $\sim$7~Mm \citep{2008gilbertApJ685} oscillating with a short period of $\sim$4 min; \cite{2014shenApJ795} analysing a higher filament ($40~\textrm{Mm}$) measured a larger period of $\sim$13 min and an amplitude of $10~$Mm; moreover, \cite{2013ApJ...773..166L} reported oscillations of an even higher filament ($63~$Mm) with an even larger period $\sim$30 min and an amplitude of ($22~$Mm). Considering the damping time, this trend is also seen in observational studies: \citet{2012ApJ...761..103G} for a prominence at a height of $36~$Mm reported a transverse oscillation period of $28~$min and a damping time of $44~$min. Following the increasing tendency, \cite{2013ApJ...773..166L} measured global oscillations in a filament of height  $63~$Mm determining a period of $30~$min and a damping time of $47~$min. This trend is also followed by the prominence studied at EUV by \cite{2011A&A...531A..53H}, who for a height of $80~$Mm reported a period of $100~$min and a damping time of $240~$min. 

As a second general rule, in both directions the results also showed that larger filaments are associated to shorter periods and smaller amplitudes. As before, this tendency partially agrees with the linear model of \citet{2016kolotkovA&A590}; taking into account that in our model an increase of the filament radius implies an  increase of the filament current $I$ (Equation \ref{e:curr}) these authors obtained that vertical periods diminish for larger filament currents $I$ but, conversely, horizontal periods increase with current $I$. On the other hand, this tendency is more difficult to compare with observations due to the lack of joint measurements of radius, period and amplitude. It seems that as larger filaments are subject to stronger net forces of higher restoring frequencies, this would imply that the filament  oscillates with shorter periods and smaller amplitudes. 

Concerning the damping mechanism obtained from the simulations, we examined the main processes discussed by several authors and found that the wave leakage seems to be the main damping mechanism.

\begin{acks}[Acknowledgements] 
We  want to thank an anonymous referee who helped us make a significant improvement to this work. EZ is grateful with FAPESP to have financed this research by the grant 2018/25177-4. MC, GK and AC are members of the Carrera del Investigador Cient\'ifico (CONICET). MC and GK acknowledge support from ANPCyT under the grant PICT No. 2016-2480. MC also acknowledge support from the SECYT-UNC grant No. 33620180101147CB. MVS thanks support from the European Research Council (ERC) under the European Union's Horizon 2020 research and innovation programme (grant agreement No 724326). GGC thanks CNPq for support with a Productivity Research Fellowship. The research leading to these results has received funding from CAPES grant 88881.310386/2018-01, FAPESP grant 2013/24155-3. EZ and GGC are also grateful with Mackenzie Research Funding Mackpesquisa for the received support. GGC is Correspondent Researcher of the Consejo Nacional de Investigaciones Cient\'ificas y T\'ecnicas (CONICET) for the Instituto de Astronom\'ia y F\'isica del Espacio (IAFE), Argentina. Simulations were ran on the IATE's clusters and CRAAM's cluster {\sc wintermute}, we thank system managers Dar\'io Gra\~na and Tiago Giorgetti. {\sc Flash} code: ``The software used in this work was developed in part by the DOE NNSA ASC- and DOE Office of Science ASCR-supported Flash Center for Computational Science at the University of Chicago''. We also thank the {\sc VisIt} team for developing the graphical tool \citep{2012Harrison2012python}.
\end{acks}


  

\bibliographystyle{spr-mp-sola}
\bibliography{./references}  

\IfFileExists{\jobname.bbl}{} {\typeout{}
\typeout{****************************************************}
\typeout{****************************************************}
\typeout{** Please run "bibtex \jobname" to obtain} \typeout{**
the bibliography and then re-run LaTeX} \typeout{** twice to fix
the references !}
\typeout{****************************************************}
\typeout{****************************************************}
\typeout{}}

\end{article} 

\end{document}